%% file: ms.tex
\documentclass[12pt,preprint]{aastex}
\shorttitle{The Central Nuclear Star Cluster}

\begin{document}

\title{High angular resolution integral-field spectroscopy of the Galaxy's nuclear cluster: a missing stellar cusp?}

\author{T. Do\altaffilmark{1}, A. M. Ghez\altaffilmark{1},  M. R. Morris\altaffilmark{1}, J. R. Lu\altaffilmark{2}, K. Matthews\altaffilmark{2}, S. Yelda\altaffilmark{1}, J. Larkin\altaffilmark{1}}
\altaffiltext{1}{Physics and Astronomy Department, University of California,
    Los Angeles, CA 90095-1547}

\altaffiltext{2}{California Institute of Technology, Pasadena, CA}

\email{tdo@astro.ucla.edu}

\begin{abstract}
We report on the structure of the nuclear star cluster in the innermost 0.16 pc of the Galaxy as measured by the number density profile of late-type giants. Using laser guide star adaptive optics in conjunction with the integral field spectrograph, OSIRIS, at the Keck II telescope, we are able to differentiate between the older, late-type ($\sim$ 1 Gyr) stars, which are presumed to be dynamically relaxed, and the unrelaxed young ($\sim$ 6 Myr) population. This distinction is crucial for testing models of stellar cusp formation in the vicinity of a black hole, as the models assume that the cusp stars are in dynamical equilibrium in the black hole potential. In the survey region, we classified 60 stars as early-type (23 newly identified) and 74 stars as late-type (61 newly identified). We find that contamination from young stars is significant, with more than twice as many young stars as old stars in our sensitivity range (K$^\prime < 15.5$) within the central arcsecond.  Based on the late-type stars alone, the surface stellar number density profile, $\Sigma(R) \propto R^{-\Gamma}$, is flat, with $\Gamma = -0.27\pm0.19$. Monte Carlo simulations of the possible de-projected volume density profile, n(r) $\propto r^{-\gamma}$, show that $\gamma$ is less than 1.0 at the 99.73 \% confidence level. These results are consistent with the nuclear star cluster having no cusp, with a core profile that is significantly flatter than predicted by most cusp formation theories, and even allows for the presence of a central hole in the stellar distribution. Of the possible dynamical interactions that can lead to the depletion of the red giants observable in this survey -- stellar collisions, mass segregation from stellar remnants, or a recent merger event -- mass segregation is the only one that can be ruled out as the dominant depletion mechanism. The lack of a stellar cusp around a supermassive black hole would have important implications for black hole growth models and inferences on the presence of a black hole based upon stellar distributions.

\end{abstract}

\section{Introduction}
Over 30 years ago, theoretical work suggested that the steady state distribution of stars may be significantly different for clusters with a massive black hole at the center than those without \citep[e.g.][]{1976ApJ...209..214B,1978ApJ...226.1087C,1980ApJ...242.1232Y}.  Stars with orbits that bring them within the tidal radius of the black hole are destroyed and their energy is transferred to the stellar cluster. In the steady state, this energy input must be balanced by the contraction of the cluster core, which appears as a steeply rising radial profile in the number density of stars toward the cluster center. This radial profile is usually characterized by a power law of the form $n(r) \propto r^{-\gamma}$, with a power law slope, $\gamma$, that is steeper than that of a flat isothermal core. For a single-mass stellar cluster, \citet{1976ApJ...209..214B} determined the dynamically relaxed cusp will have $\gamma = 7/4$. The presence of such a steep core profile, or cusp, is important observationally because it may represent a simple test for black holes in stellar systems where dynamical mass estimates are difficult, such as in the cores of galaxies. The stellar cusp is also important theoretically as it is a probable source of fuel for the growth of supermassive black holes and its presence is often assumed in simulations of stellar clusters having a central black hole \citep[e.g.,][and references therein]{2005gbha.conf..221M}.

Theoretical work has progressed from the early simulations by \citet{1976ApJ...209..214B} to include many complicating effects, including multiple masses, mass segregation, and stellar collisions, on the density profile of stellar clusters in the presence of a supermassive black hole \citep[e.g.,][]{1977ApJ...216..883B,1991ApJ...370...60M,2008arXiv0808.3150A}. These theories predict cusp slopes within a black hole's gravitational sphere of influence ranging from $ 7/4 \geq \gamma \geq 7/3$ for a multiple mass population to as shallow as $\gamma = 1/2$ for a collisionally dominated cluster core. An important assumption of all cusp formation models is that the stellar cluster be dynamically relaxed. Without this assumption, the stellar distribution would show traces of the cluster origin in addition to the influence of the black hole.

The Galactic center is an ideal place to test these theories of cusp formation as it contains the nearest example of a supermassive black hole \citep[Sgr A*, with a mass M$_{\bullet} = 4.1\pm0.6\times10^{6}$ M$_{\odot}$,][]{2008ApJ...689.1044G,2009ApJ...692.1075G}. Located at a distance of only 8 kpc, the radius of the sphere of influence of the Galactic center black hole ($\sim1$ pc) has an angular scale of $\sim 25\arcsec$ in the plane of the sky, two orders of magnitude larger than any other supermassive black hole.  In order to plausibly test stellar cusp formation theories, the stellar population used to trace the cusp profile must be older than the relaxation time, which is of order 1 Gyr within the sphere of influence of Sgr A* \citep[e.g.,][]{2006ApJ...645.1152H}. In the Galactic center, late-type red giants (K to M) are the most promising tracers of the stellar distribution since they are $> 1$ Gyr old, bright in the NIR, and abundant. On parsec size scales, these stars dominate the flux, so early seeing limited observations of the surface brightness profile in the NIR at the Galactic center were successfully used to show that the structure of the nuclear star cluster at large scales has a density power law of $\gamma \approx 2$ \citep{1968ApJ...151..145B}. Integrated light spectroscopy of the CO absorption band-head at 2.4 $\micron$ (which is dominated by red giants) also confirmed this slope down to within $\sim 0.5$ pc from the black hole \citep{1989ApJ...338..824M,1996ApJ...456..194H}. These measurements, as well as integrated light spectroscopy from \citet{2000ApJ...533L..49F} of the inner 0\arcsec.3, found a lack of CO absorption in the inner $\sim 0.5$ pc of the Galaxy. It was unclear at the time whether this represented a change in the stellar population with fewer late-type stars in the inner region or a change in the stellar density profile. Because the integrated light spectrum is biased toward the brightest stars, contamination from a few young Wolf-Rayet (WR) stars (age $< 6$ Myr), such as the IRS 16 sources identified in the early 1990s located between 1--2$\arcsec$ from the black hole \citep[e.g.,][]{1990MNRAS.244..706A,1991ApJ...382L..19K}, can significantly impact measurements of the underlying stellar density.

Several approaches have been taken to identify the cause of the difference in the CO equivalent width in the integrated light within the inner 0.5 pc of the Galaxy. \citet{2003ApJ...594..294S} attempted to remove the effects of the bright young stars on the surface brightness distribution through narrow band imaging with NICMOS/HST. They found a surface brightness that began to flatten and drop slightly in the central 0.1 pc. Since fainter young stars can still contribute to the surface brightness, \citet{2003ApJ...594..812G} used adaptive optics (AO) imaging with $H$ (1.6 \micron) and $K_{s}$ (2 \micron) filters to measure the stellar density using number counts, which they argue more robustly traced the population of red giants. They find $\gamma = 2.0\pm0.1$ at $R \ge 10\arcsec$ and $\gamma = 1.4\pm0.1$ at $R < 10\arcsec$, which suggests that there may be a stellar cusp at the Galactic center within the range predicted by \citet{1977ApJ...216..883B}. Further work by \citet{2007A&A...469..125S}, which included deeper AO imaging, produced a fit of a broken power law to the observed projected surface number density profile of the form $\Sigma(R) \propto R^{-\Gamma}$, where $R$ is the projected distance from Sgr A* in the plane of the sky, and $\Gamma$ the projected power law slope. Their best fit broken power law model has a break radius at $R_{break} = 6\arcsec.0\pm1\arcsec.0$ (0.22 pc), with $\Gamma = 0.75\pm0.1$ outside $R_{break}$ and  $\Gamma = 0.19\pm0.05$ as the slope of the stellar cusp inside $R_{break}$. This indicated that the cusp slope is perhaps flatter than the range predicted by \citet{1977ApJ...216..883B}. However, with only number counts, it was still unclear how much bias to the slope was contributed by the dynamically unrelaxed young stellar population, which represents a large fraction of the bright stars ($K^{\prime} < 14$) that have been spectroscopically identified in the inner $~1\arcsec$ \citep[e.g.,][]{2003ApJ...586L.127G,2006ApJ...643.1011P}. 

To make further progress, it is necessary to discriminate between the early and late-type stars. One recent approach has been through narrow-band imaging to target wavelengths where there are measurable differences between spectral features of the two types of stars. For example, the CO bandhead absorption features starting at 2.29 $\micron$ are very strong for late-type stars (K \& M), and absent for early-type stars (O \& early B). Using the CO narrowband images from the Gemini North Galactic Center Demonstration Science Data Set, \citet{2003ApJ...594..812G} found that the fraction of late-type stars declines between $10\arcsec$ and $1\arcsec$ from the Galactic center, and thus, may be flatter than expected from number counts. However, this narrowband dataset was shallower than that of the broadband images used in the number counts measurements and had trouble differentiating between the two types of stars at $K > 13$. More recently, \citet{2009arXiv0903.2135B} took deeper narrowband images ($K < 15.5$)  in CO and found an even more shallow slope in the inner region of the cluster. Their best fit slope of $\Gamma = -0.17\pm0.09$ at $R < 6\arcsec.0$ suggests that the late-type stars may even be slightly declining toward the cluster center. While number counting with narrowband images has greater discriminating power than broadband images, this method has limitations. First, this method is strongly dependent on how well the dataset can be calibrated based upon existing spectroscopically determined samples, which is limited to K$< 13$. In addition, limits on the CO equivalent widths depend on how precisely the slope of the stellar continuum can be measured with narrow band filters. 

A spectroscopically selected sample would be ideal to separate the late-type, presumably dynamically relaxed population from that of the young early-type stars. Until recently, much of the spectroscopy taken of stars in the Galactic center has been seeing-limited and limited in sensitivity to $K < 13$ \citep[see][]{2000MNRAS.317..348G,2003ApJ...597..323B,2003ApJ...599.1139F}. These surveys found a drop in the fraction of late-type stars toward the central $\sim10\arcsec$, but because they were sensitive only to the brightest giants, the uncertainty remained as to whether the fainter stars would show the same trend. Nevertheless, these results combined with measurements from imaging all point toward fewer stars in the central 0.5 pc than would be expected for a relaxed cusp. 

By combining the advantages of high spatial resolution from AO imaging with the discriminating power of spectroscopy, the advent of integral-field spectroscopy (IFU) behind AO has potential to provide the most sensitive method of measuring the stellar cusp profile in the inner 0.5 pc of the Galaxy. Measurements with the SINFONI IFU at the VLT have shown that it is possible to identify young stars at better than 100 mas angular resolution \citep{2006ApJ...643.1011P,2009ApJ...692.1075G}. Here, we report on a high angular resolution spectroscopic survey of the central 0.15 parsec of the Galaxy using laser guide star adaptive optics with the OSIRIS IFU at the Keck II telescope. The survey reaches a completeness of 40\% at $K^{\prime} \approx 15.5$, two magnitudes fainter than previously reported spectroscopic surveys of late-type stars in this region. The observations and data reduction method are described in Section \ref{sec:obs} and \ref{sec:data}. Using these observations, it is possible to separate the presumably dynamically relaxed late-type population from the young, unrelaxed population (Section \ref{sec:spec}). We find that the surface number density of late-type stars is significantly flatter than the range of power laws predicted by \citet{1977ApJ...216..883B} (Section \ref{sec:results}). This measurement rules out all values of $\gamma > 1.0$ at a confidence level of 99.73\%. In Section \ref{sec:discussion}, we discuss possible dynamical effects that may lead to the flat observed slope and implications for future measurements of the stellar cusp at the Galactic center. 

\section{Observations}
\label{sec:obs}

Near-IR spectra of the central 4$\arcsec$ of the Galaxy were obtained between 2006 and 2008 using the OSIRIS integral field spectrograph \citep{2006NewAR..50..362L} in conjunction with the laser guide star adaptive optics (LGS AO) system on the Keck II telescope \citep{2006PASP..118..297W, 2006PASP..118..310V}. The laser guide star was propagated at the center of our field and for low-order tip-tilt corrections, we used the $R = 13.7$ mag star, USNO 0600-28577051, which is located $\sim19\arcsec$ from Sgr A*. The AO system enabled us to obtain nearly diffraction limited spatial resolution ($\sim$ 70 mas) in our long-exposure (15 min) 2 $\micron$ data sets. All observations were taken with the narrow band filter Kn3 (2.121 to 2.229 $\micron$) centered on the Br $\gamma$ hydrogen line at 2.1661 $\micron$, with a pixel scale of 35 mas and a spectral resolution of $\sim3000$. With this filter, OSIRIS has a field of view of $1.\arcsec 68 \times 2.\arcsec 24$ per pointing and offers sensitivity to both early and late-type stars. Early-type stars from OB supergiants to B type main sequence young stars show a prominent Br $\gamma$ line, while late-type (K \& M) evolved giants contain strong \ion{Na}{1} lines in this wavelength range.

\begin{figure}
\centering
\includegraphics[width=5in]{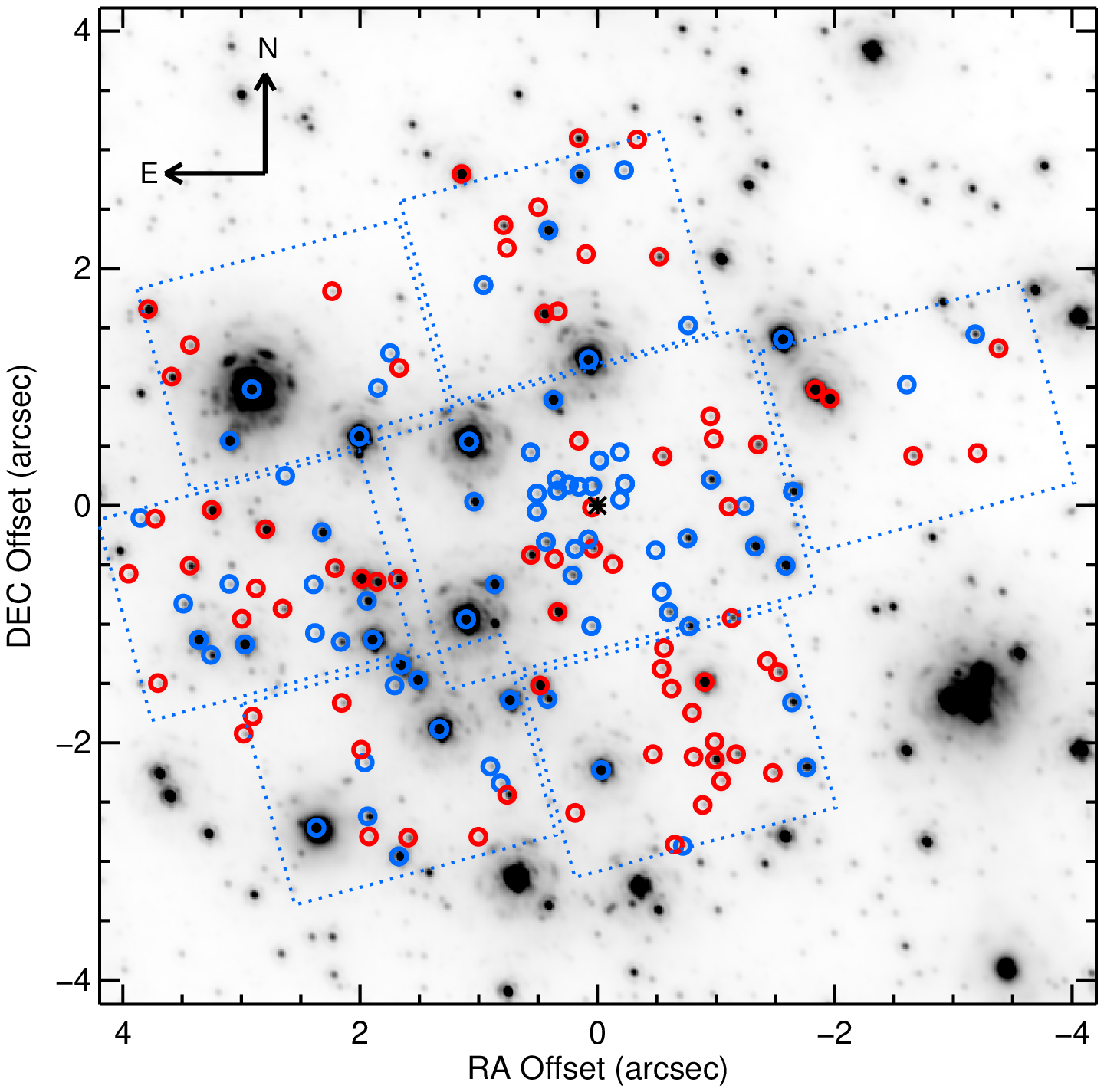}
\caption{The currently surveyed region overlaid on a K$^\prime$ (2.2 \micron) image of the Galactic center taken in 2007. Sgr A* is marked at the center with a *. Spectroscopically identified early (blue) and late-type (red) stars are marked with circles. Each field is enclosed by dotted lines. Some spectral identifications are outside of the marked lines because they were found at the edge of the dithers.}
\label{fig:mosaic_fov}
\end{figure}

With the goal of covering an $\sim 8\arcsec\times6\arcsec$ area centered on Sgr A*, we divided this region into 9 fields (see Figure \ref{fig:mosaic_fov}). Seven of the 9 fields have now been observed and we will refer to these pointings by their approximate orientation with respect to Sgr A* (i.e. GC Central, GC East, GC Southeast, GC South, GC West, GC North, and GC Northeast).  We obtained roughly ten 900 sec exposures ($\sim 2.5$ hr total exposure time) for each field, with the exception of the Northeast field, for which we obtained five exposures. For all but the central field, the exposures were dithered in a nine position pattern that covered the center, sides and corners of a $0\arcsec.2\times0\arcsec.2$ square. For the field centered on Sgr A*, the dither positions were the four corners of a $1\arcsec \times 1 \arcsec$ square in an effort to keep the $1\arcsec \times 1 \arcsec$ surrounding Sgr A* within the field of view of all exposures. The survey region is oriented at a position angle (PA) of $\sim105$ degrees, with the longer direction oriented along major axis of the disk of young stars rotating in the clockwise direction in the plane of the sky \citep{2006ApJ...643.1011P,2009ApJ...690.1463L}. Table \ref{table:obs} summarizes these observations.

\section{Data Reduction}
\label{sec:data}

Data cubes with two spatial dimensions and one spectral dimension were produced using the OSIRIS data reduction pipeline provided by Keck Observatory. The pipeline performs dark current subtraction, cosmic ray removal, and wavelength rectification. The cubes were not corrected for the effects of chromatic differential atmospheric dispersion because this effect is small within the wavelength range of Kn3, shifting the centroid of stars by only 0.1 spatial pixel between the shortest and longest wavelength channel. 

For each field, stellar positions were estimated by running \textit{StarFinder}, a point spread function (PSF) fitting program \citep{2000A&AS..147..335D}, on images made from collapsing the data cubes along the spectral dimension. First, a mosaic was made by combining the individual dithers within each field with the OSIRIS pipeline using an average for all overlapping spatial pixels. An image was then created from the mosaicked cube by taking the median of the spectral channels at each spatial pixel. Next, between 1--3 of the brightest, most isolated stars were chosen by hand as PSF reference stars for \textit{StarFinder}. This PSF was then used to determine the centroids of all the stars in the mosaic in an initial pass. After the first attempt at identifying sources, we ran an additional iteration of \textit{StarFinder} to improve the estimate of the PSF, which is then used in a second pass to determine centroids. The stars identified by this process were cross-checked with the nearest epoch of deep images taken at $K^\prime$ using LGS AO with the NIRC2 imager on Keck II \citep{2008ApJ...689.1044G} in order to remove false detections in the noisier OSIRIS data and to obtain accurate broadband photometric measurements for these stars. As a final step, \textit{StarFinder} was run on each individual cube using relative positional priors from the deeper mosaic cubes and the PSF as determined from the OSIRIS mosaic to obtain precise OSIRIS positions in preparation for spectral extraction from the individual data cubes.

The spectrum of every star was then extracted from each cube by performing aperture photometry for the star at each spectral channel. We used an aperture of radius 2 pixels (70 mas) to extract the stellar spectrum and the median of an annulus with inner radius at 2 pixels and outer radius at 4 pixels to determine the sky background. For regions with very high stellar density, such as within the central arcsecond, the background was manually determined by the median of patches of sky free of stars as close to the target star as possible. The final spectrum of each star was then produced by averaging the spectra of that star from all data cubes from that night. When the signal to noise ratio (SNR) per spectral channel between the spectra were significantly different, they were combined using an average weighted by the SNR (calculated between 2.212 to 2.218 \micron). 

The combined spectra have SNRs ranging from over 100 for bright stars (K$^\prime < 13$) to $\sim 5$ for the faint stars (K$^\prime \approx 15.5$) that have marginal line detections. Photon noise appears to dominate the SNR of the final spectra and thus our ability to assign spectral types to most of the stars. In addition, regions with strong gas emission such as along the mini-spiral \citep{2004A&A...426...81P} and within the central arcsecond can be problematic for identifying Br $\gamma$ absorption lines from early-type stars. Interstellar Br $\gamma$ emission can be quite strong and can have several kinematic components, which can lead to a systematic bias in the measurement of the Br $\gamma$ line in the early type stars if the background gas component is not completely corrected. Confusion is also a problem around very bright sources such as the IRS 16 stars, where we were unable to extract several $K^{\prime} \approx 13$ stars that were identified in NIRC2 images\footnote{NIRC2 is less sensitive to confusion because the plate scale is 10 mas compared to 35 mas used in the OSIRIS observations and the higher strehl ratios obtained from the much shorter exposures (30 s vs. 900 s)}. 

An A0V (HD 195500) and G2V (HD 193193) star were observed on each night to correct for atmospheric absorption lines\footnote{These stars were observed using natural guide star mode with 30 s integration time and 5 frames per star.}. The A star, having no intrinsic absorption lines other than Br $\gamma$ and He I in the Kn3 filter, serves as a good indicator of atmospheric and instrumental features. To remove the Br $\gamma$ and He I line from the A0V spectrum, we replace the region between 2.155 to 2.175 $\micron$ with the same region from the G2V star and divided by the empirically observed solar spectrum shifted to the radial velocity of the G2V star (similar to the procedure described in \citet{1996ApJS..107..281H}). The A star spectrum was then divided by a 10,000 K blackbody to remove the stellar continuum.

\section{Spectral Identification}
\label{sec:spec}

Given our survey sensitivity, we expect to be able to detect the early-type stars in the form of bright Wolf-Rayet stars, OB supergiants, main sequence young stars down to B2-B4V, and late-type stars in the form of K-MIII evolved giants. The most important feature of this survey is its ability to differentiate between early and late-type stars. This distinction is apparent when comparing template spectra of early type stars with those of late type giants, even in the limited Kn3 wavelength range; the young stars have featureless spectra except for Br $\gamma$ and He I at 2.1661 and 2.1641 \micron, respectively, while the late-type stars lack these high temperature lines and show a plethora of other atomic and molecular lines, most prominently the \ion{Na}{1} lines at 2.2062 and 2.2090 \micron. 

\begin{figure}
\centering
\includegraphics[width=5in]{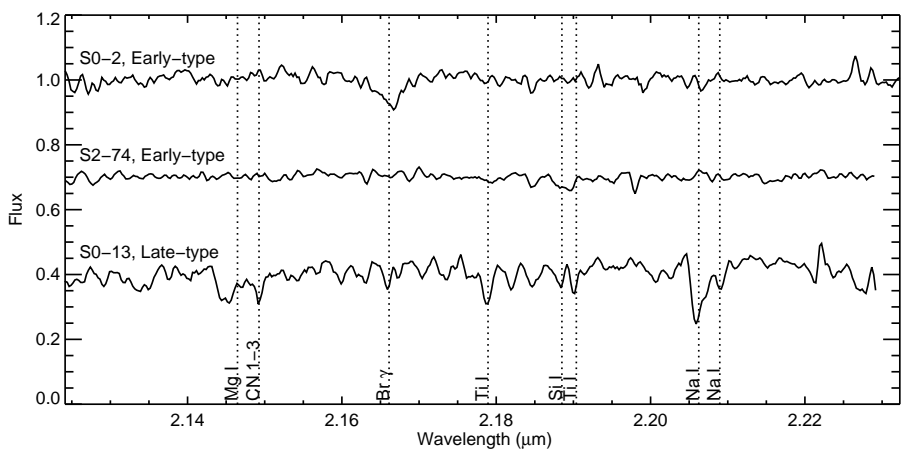}
\caption{Example of observed spectra in the Kn3 narrowband wavelength region of early and late-type stars. The Br $\gamma$ and \ion{Na}{1} lines are used to differentiate between early and late-type stars, respectively. S0-2 is a $K^{\prime} = 14.1$ early-type star with a strong Br $\gamma$ line. S2-74 is a  $K^{\prime} = 13.3$ early-type star showing a featureless spectrum in the OSIRIS Kn3 filter wavelength range. S0-13 is a  $K^{\prime} = 13.5$ late-type star showing the \ion{Na}{1} doublet. See Figures 2a, 2b, and 2c in the online supplements for spectra of all stars with spectral types identified in this survey.}
\label{fig:spectra}
\end{figure}

\begin{figure}
\centering
\includegraphics[width=3.5in,angle=90]{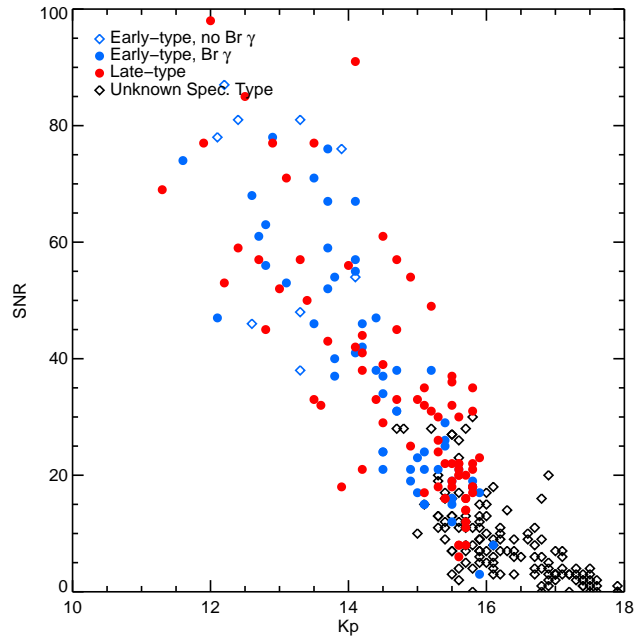}
\caption{The relationship between the SNR and $K^{\prime}$ for stars observed in this study and their spectral type. The spectral types of every star with SNR $> 40$ was determined successfully.}
\label{fig:spec}
\end{figure}

In order to assign spectral types to the stars, we separate all the extracted spectra by visual inspection into five groups: 1. Spectra with prominent Br $\gamma$ lines and no \ion{Na}{1} lines are identified as early-type stars (e.g., S0-2 in Figure \ref{fig:spectra}). 2. Spectra with prominent \ion{Na}{1} lines, but no Br $\gamma$ line are identified as late-type stars (e.g., S0-13 in Figure \ref{fig:spectra}). 3. Low SNR ($ <5$) spectra with no apparent lines are classified as unknown. 4. Bright stars with high SNR (K$^\prime < 14$, SNR $> 40$) spectra with no late-type or Br $\gamma$ lines are identified as young. These are most likely late O-type stars, which have very small or undetectable Br $\gamma$ absorption lines at the resolution of this survey \citep{1996ApJS..107..281H} (e.g., S2-74 in Figure \ref{fig:spectra}). 5. Spectra with both Br $\gamma$ and \ion{Na}{1} lines are identified as late-type if there is strong gas emission nearby, which would explain the Br $\gamma$ line as an over-subtraction of the background gas (2 stars showed these features). In the survey region, we classified 60 stars as early-type (23 newly identified) and 74 stars as late-type (61 newly identified). The late and early-type stars are reported in Tables \ref{table:late} and \ref{table:early} and labeled in Figure \ref{fig:mosaic_fov}. The stars with successfully extracted spectra, but have undetermined spectral types are listed in Table \ref{table:unknown}. Our ability to classify the stars is strongly dependent on the SNR of the spectra, with 100\% of the stars having SNR $> 40$ being identified (see Figure \ref{fig:spec}). 

\begin{figure}
\centering
\includegraphics[angle=90,width=5in]{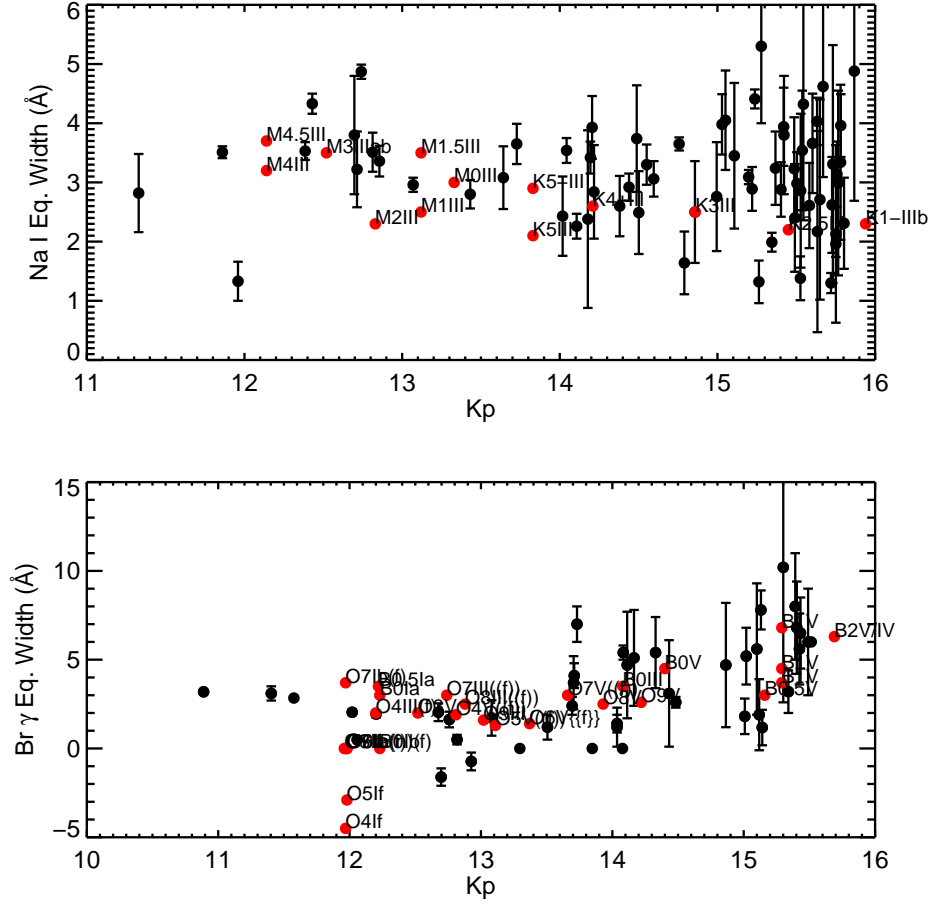}
\caption{\textbf{Top}: equivalent width of the combination of \ion{Na}{1} lines at 2.2062 and 2.2090 $\micron$ as a function of $K^\prime$ magnitude. Also plotted are equivalent width measurements of \ion{Na}{1} from \citet{2000AJ....120.2089F} with spectral types corrected for the distance and extinction (A$_k$=3.0) toward the Galactic center (red).  \textbf{Bottom:} equivalent width of the Br $\gamma$ hydrogen line at 2.1661 \micron. Equivalent width measurements from \citet{1996ApJS..107..281H} are also plotted for O and B type stars, corrected for the distance and extinction (A$_k$=3.0) to the Galactic center (red).}
\label{fig:ew}
\end{figure}

Equivalent widths for the \ion{Na}{1} lines in the spectra of the late-type stars were measured by normalizing the spectra and integrating between 2.2053 to 2.2101 $\micron$ as in \citet{2000AJ....120.2089F}, after shifting the spectra to rest wavelengths. The \ion{Na}{1} equivalent widths range from 1.2 to 4.9 \AA, consistent with that of KIII to MIII evolved giants, with RMS uncertainties between 0.1 to 2 \AA. There does not appear to be a significant correlation between the equivalent width and $K^\prime$ magnitude, suggesting that the \ion{Na}{1} lines are not able to further differentiate between K and M spectral types at the SNR of these measurements (see Figure \ref{fig:ew}). The equivalent widths of the Br $\gamma$ lines are measured by fitting a Gaussian to the line and integrating the model Gaussian parameters. While the equivalent widths of the \ion{Na}{1} lines do not change with brightness, the young stars show a significant decrease in the equivalent width of Br $\gamma$ with increasing brightness as shown in Figure \ref{fig:ew}. These measurements support the assumption that the bright stars ($K^\prime < 13$) with no spectral features in the Kn3 wavelength range are young stars. 

\begin{figure}
\centering
\includegraphics[width=5in]{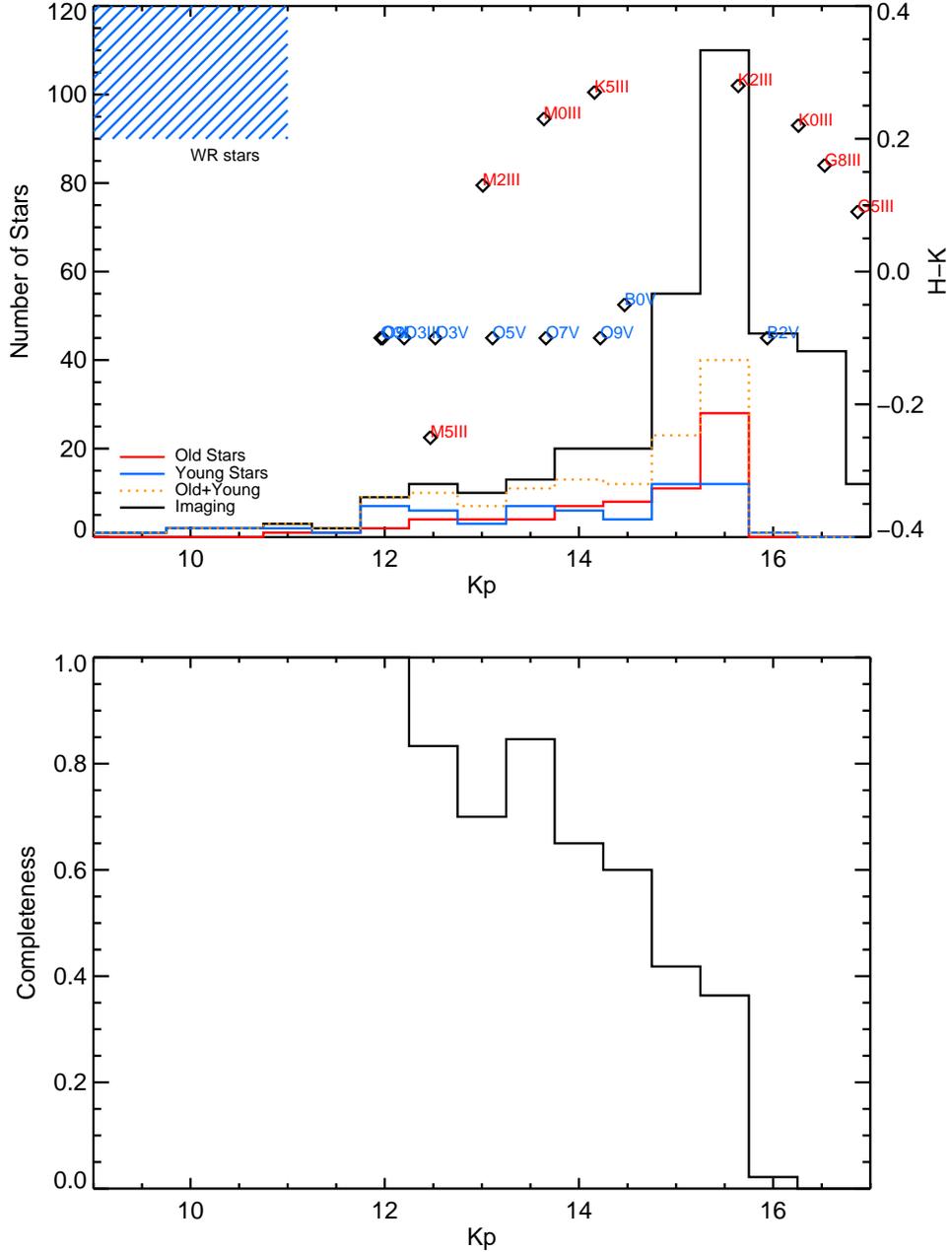}
\caption{The $K^\prime$ luminosity function from the spectroscopic survey in comparison to that found from imaging. In the top plot, we also include for reference, the rough spectral types expected to be observable at each luminosity bin assuming $A_{K} = 3.0$ and a distance of 8 kpc. The axis on the right shows the H-K color associated with each of the spectral types \citep{2001ApJ...558..309D, 2006A&A...457..637M}. The WR stars occupy a range in both $K^{\prime}$ and H-K colors.}
\label{fig:kluminosity}
\end{figure}

\begin{figure}
\centering
\includegraphics[angle=90,width=6.5in]{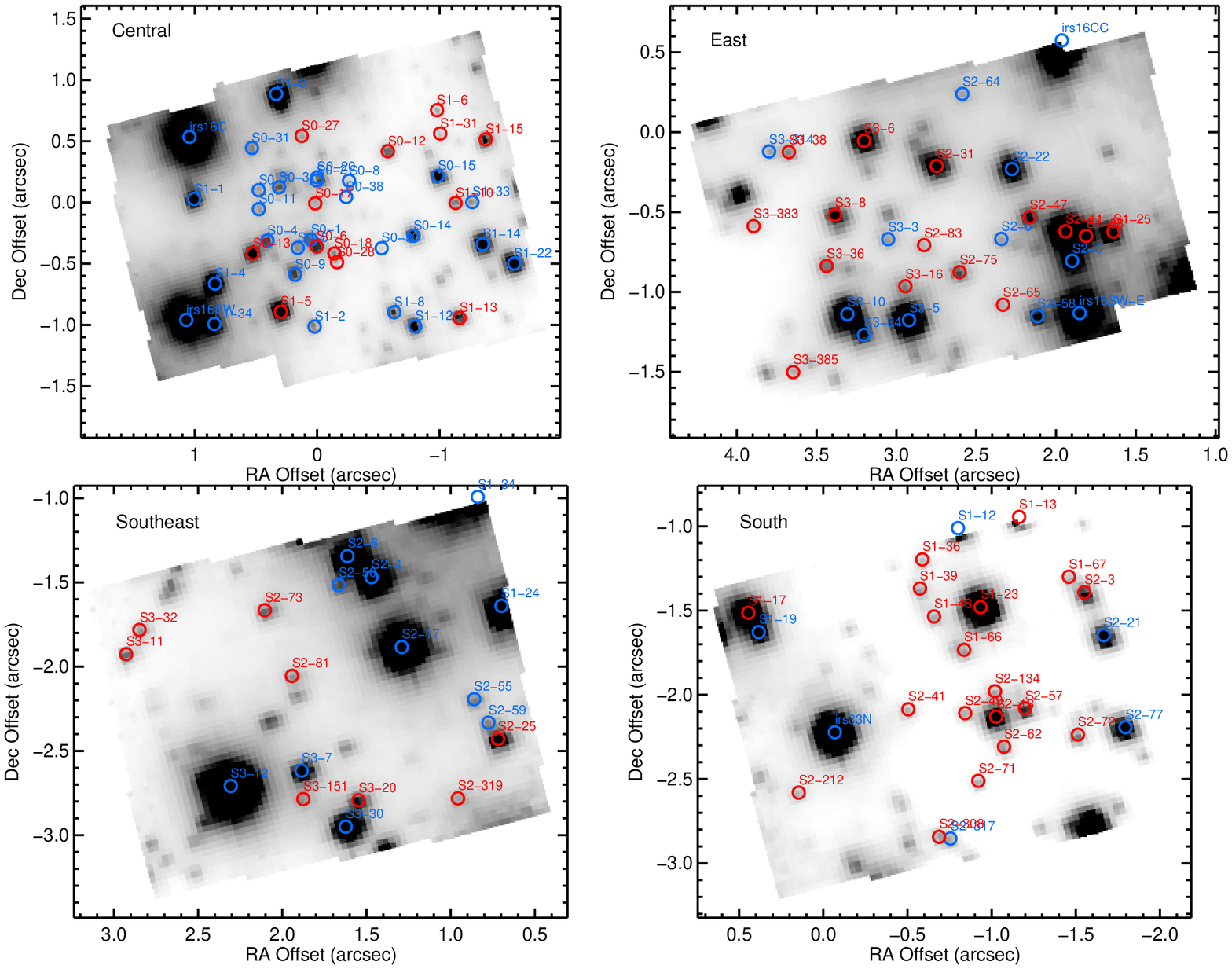}
\caption{Images from collapsing the OSIRIS data cubes along the spectral dimension for each individual pointing of the survey. The images are oriented with north up and east to the left. Spectroscopically identified early (blue) and late-type (red) stars are marked with circles.}
\label{fig:osiris_fields1}
\end{figure}

\begin{figure}
\centering
\includegraphics[angle=90,width=6.5in]{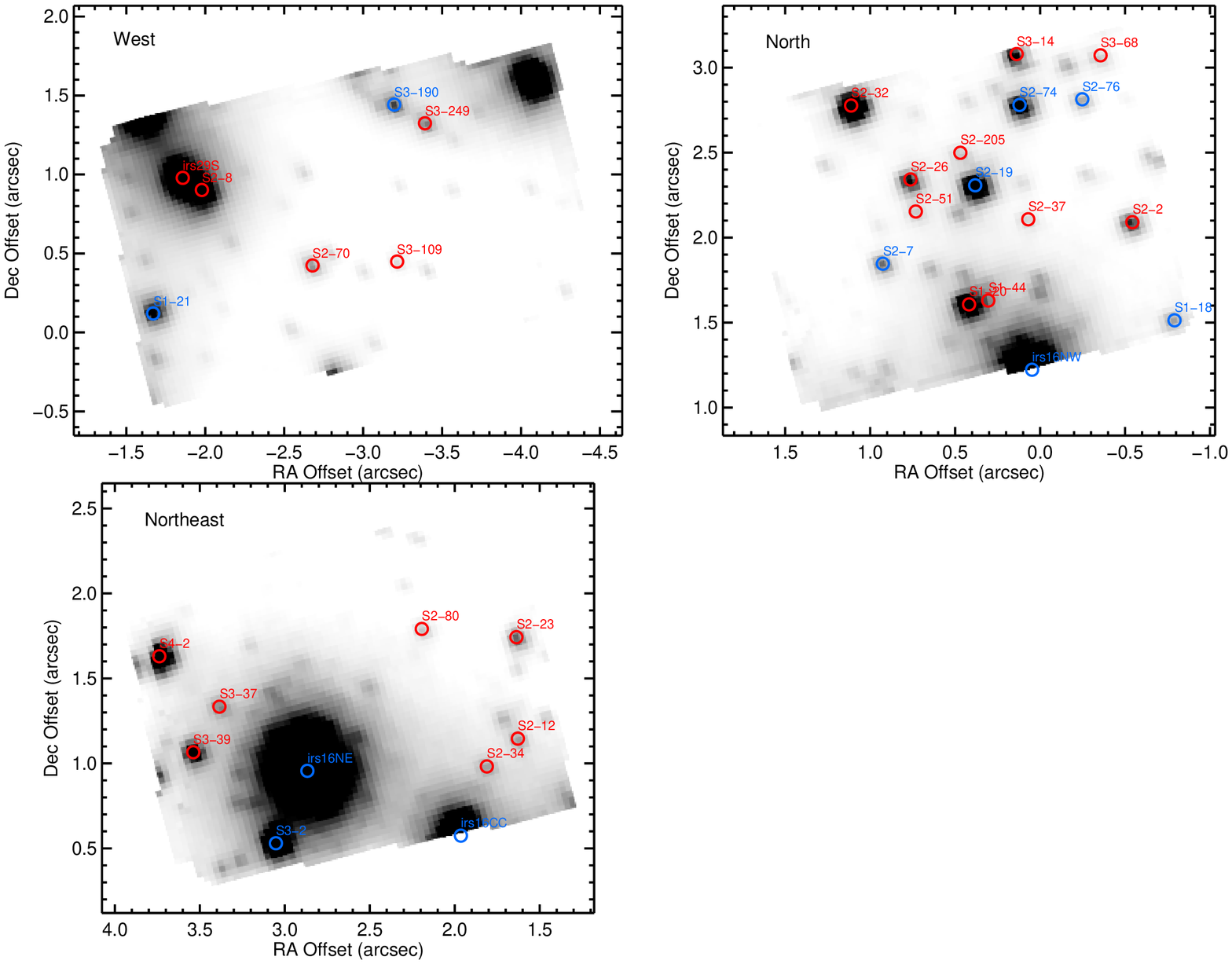}
\caption{Images from collapsing the OSIRIS data cubes along the spectral dimension for each individual pointing of the survey. The images are oriented with north up and east to the left. Spectroscopically identified early (blue) and late-type (red) stars are marked with circles.}
\label{fig:osiris_fields2}
\end{figure}

This survey is complete to about $K^\prime$ = 12 and falls off to about 40\% completeness at $K^\prime = 15.5$, measured relative to the stars detected in the same region from imaging, which is complete to $K^\prime = 16.0$ \citep{2008ApJ...689.1044G}. We were able to determine spectral types for all stars with extractable spectra and $K^{\prime} < 14.0$. The sample is not complete to $K^{\prime} = 14.0$ because the spectra of some bright stars ($K^{\prime}\approx 13$) detected in broadband images were not extractable due to confusion with the IRS 16 sources. The $K^\prime$ luminosity function of all the stars in the survey is shown in Figure \ref{fig:kluminosity}, with a comparison to that derived from imaging at $K^\prime$. The $K^{\prime}$ luminosity functions of the stars in the individual fields are similar to that of the total sample, with the notable exception of the western field, where there are fewer stars detected both in imaging and in spectroscopy. This difference in number of detected stars is most likely due to poor seeing on the night of observation and higher extinction in that region. Figures \ref{fig:osiris_fields1} and \ref{fig:osiris_fields2} shows the locations of the stars with their spectral types on a collapsed image of each of the OSIRIS fields. 

\section{Results}
\label{sec:results}
\begin{figure}
\centering
\includegraphics[width=3.5in]{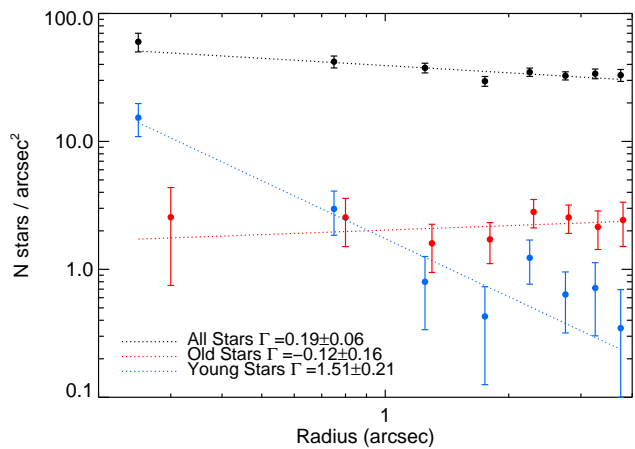}
\caption{Plot of the surface number density as a function of projected distance from Sgr A* in the plane of the sky for different populations: old (late-type, red), young (early-type, blue), and total number counts from $K^\prime$ imaging.}
\label{fig:radial}
\end{figure}

With spectroscopic identification of the stars brighter than $K^\prime \sim 15.5$, we are able to separate the presumably dynamically relaxed old stars from the unrelaxed young population. The stellar number density profile is constructed in radial bins of 0.\arcsec5 over the area that was sampled in this survey out to a radius of $\sim 3\arcsec$, with error bars scaling as $\sqrt{N}$, where $N$ is the number of stars in each bin. It is clear that, while the number density of early-type stars increases quite rapidly toward the central arcsecond, the late-type stars have a very flat distribution (Figure \ref{fig:radial}). Outside the central arcsecond, the projected number density of early-type stars drops to about half that of the late-type stars. For a quantitative comparison to stellar cusp models, we fit a power law to the surface number density of both populations. The late-type population is best fit with a power law slope $\Gamma_{old} = -0.12 \pm 0.16$, while for the early-type stars, $\Gamma_{young} = 1.51\pm0.21$. For comparison, we also measured the projected stellar number density over the same survey region using LGS AO $K^\prime$ imaging from July 2008, which is much deeper ($K^\prime < 20$), but with no spectroscopic discrimination. We find that $\Gamma_{imaging} = 0.19 \pm 0.06$, comparable to that found by \citet{2007A&A...469..125S} using only number counts. The spectroscopically determined late-type slope is flatter than that determined from number counts alone, but is consistent within the $\sim 2 \sigma$ formal fitting errors. 

The survey has complete azimuthal coverage out to a radial distance of $\sim 1.\arcsec5$ from Sgr A*. Beyond this radius, the radial coverage extends out to $\sim4\arcsec$, but the azimuthal coverage is complete due to two fields missing in the northwest and southwest. We find that these missing two fields do not significantly impact the radial profile of the late-type stars, as removing one or two of the outer observed fields from the analysis leads to the same power law slope, within the measurement uncertainties. In addition, previous observations of the late-type stars in this region have found the distribution to be isotropic \citep[e.g.][]{2003ApJ...594..812G,2008A&A...492..419T}.

\begin{figure}
\centering
\includegraphics[width=3.5in]{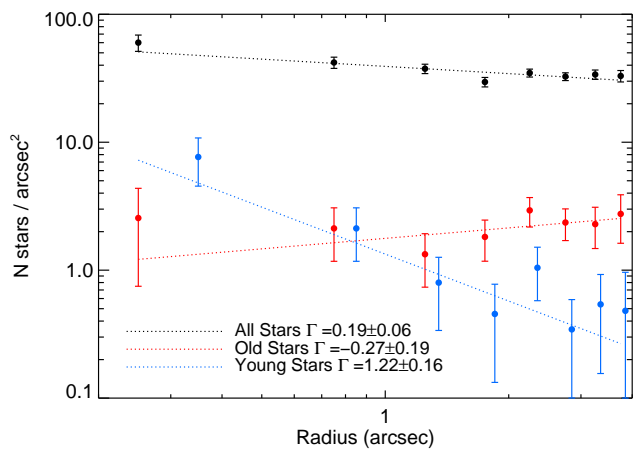}
\caption{Plot of the surface number density as a function of projected distance from Sgr A* in the plane of the sky for different populations: old (late-type, red), young (early-type, blue), and total number counts from $K^\prime$ imaging. These radial profiles have been corrected for completeness and extinction using the method detailed in Section \ref{sec:results}.}
\label{fig:radial_corr}
\end{figure}

Potential systematic errors to the radial profile measurement may arise from variable extinction and incompleteness between each field. In order to study how these two factors impact our result, the radial profile was computed using stars with extinction corrected magnitudes only over regions that are at least 30\% complete. To do so, the extinction map reported by \citet{2009arXiv0903.2135B} was used to correct for the variations in extinction between each star by adding $A_K - 3.0$ to each star to bring them to the same canonical $A_K = 3.0$ extinction to the Galactic center (see Table \ref{tab:ext} for the average and RMS value of $A_{K}$ in each field). Then, the $K^\prime$ luminosity function was recomputed for each field to determine the $K^\prime$ magnitude bin beyond which the completeness falls to less than 30\%. Table \ref{tab:ext} shows this magnitude bin for each field before and after extinction correction. This separates the fields into three groups in terms of their 30\% completeness : all fields are at least 30\% complete down to $K^\prime = 14.5$; all fields except W are complete to $K^\prime = 15.5$; the central field is complete to $K^\prime = 16.0$. We calculate the radial profiles in each magnitude bin using only fields that are at least 30\% complete at that magnitude. For all stars with $K^\prime < 14.5$, we use all the fields, while for stars with $14.5 \le K^\prime < 15.5$, the western field was dropped from the measurement; the $15.5 \le K^\prime < 16.0$ magnitude bin was not included because only the central field is complete to at least 30\% at that magnitude. The radial profiles from the $K^\prime < 14.5$ and $14.5 \le K^\prime < 15.5$  magnitude bins were then summed to produce the final measurement of the surface density, with errors added in quadrature (see Figure \ref{fig:radial_corr}). A total of 32 early-type and 63 late-type stars are included in the final radial profile measurement. The early-type stars have $\Gamma_{young} = 1.22\pm0.16$, while for the late-type stars, $\Gamma_{old} = -0.27\pm0.19$. Neither of these values differ significantly from the case without extinction and completeness correction. The rest of the analysis will use these values for the measured surface number density profile. 

Beyond a radius of 1\arcsec, there are more young stars detected in the eastern and southeastern fields compared to the other observed regions. This excess is in the direction of the major axis of the projected coherent disk of young stars orbiting in the clockwise direction in the plane of the sky \citep{2009ApJ...690.1463L}. however, after accounting for the completeness and differences in $A_{K}$, neither the number of young, nor old stars shows a statistically significant excess. 

\section{Discussion}
\label{sec:discussion}

\begin{figure}
\centering
\includegraphics[width=2.5in,angle=90]{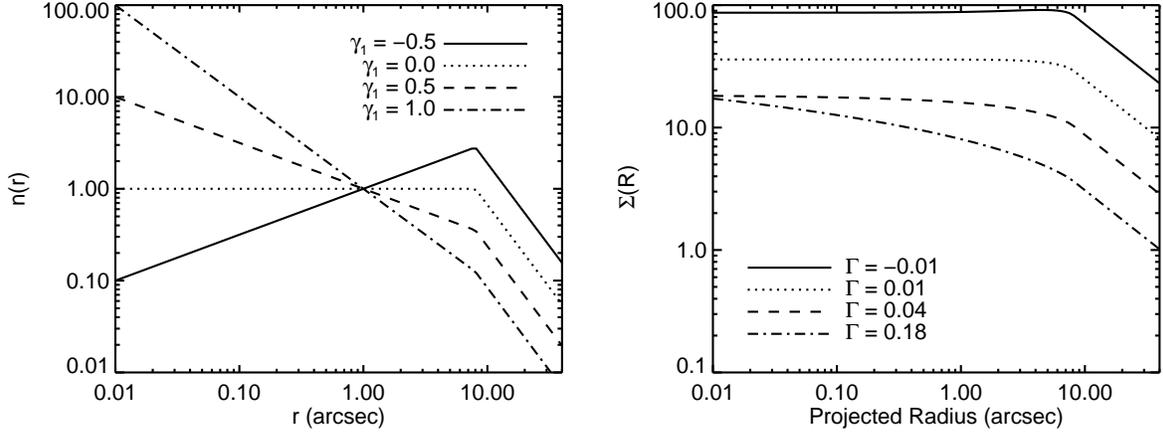}
\caption{Left: broken power law density profiles with break radius, $r_{break} = 8.0\arcsec$ and outer power law $\gamma_{2} = 2.0$, and varying inner power laws $\gamma_{1}$. Right: the projected surface number density profile of each of the broken power laws. The fitted inner surface density power law $\Gamma$ is flat for $\gamma_{1} \lesssim 0.5$.}
\label{fig:deproject}
\end{figure}

\begin{figure}
\centering
\includegraphics[width=3in,angle=90]{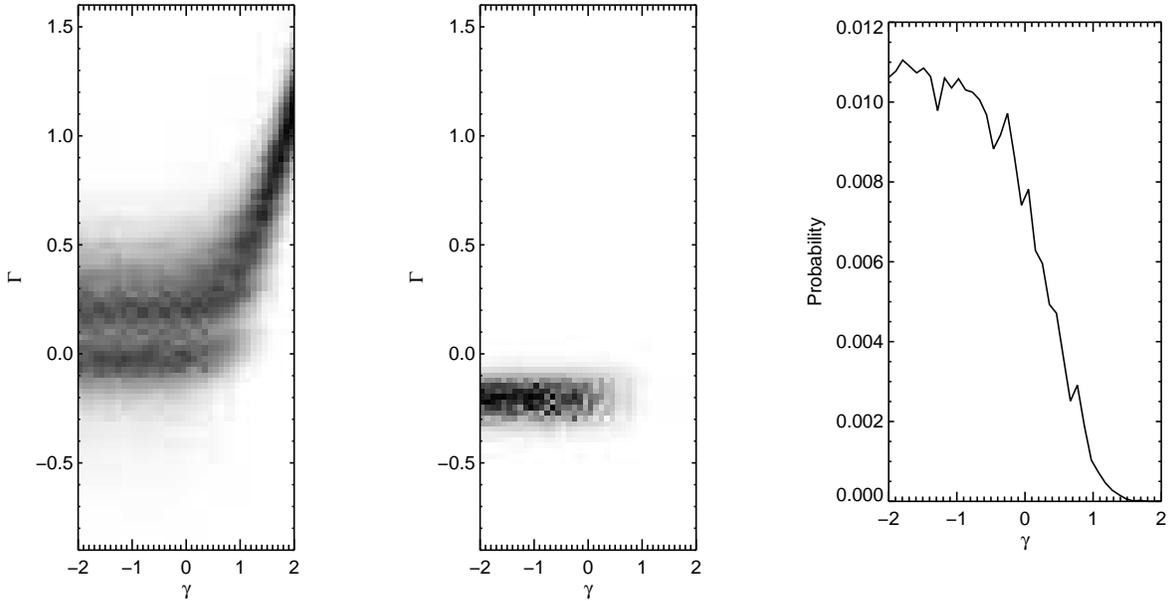}
\caption{Left: the results of the Monte Carlo simulations of different inner radial power laws $\gamma$ and the resulting measured projected power law of the radial profile, $\Gamma = -0.27\pm0.19$. Center: using our measurements, we can constrain some of the parameter space for $\Gamma$ vs. $\gamma$. Right: we marginalize over $\Gamma$ to determine that our measurement constrains $\gamma$ to be less than 1.0 at a 99.73\% confidence level.}
\label{fig:simcluster}
\end{figure}

The effects of projection of a three-dimensional cluster onto the plane of the sky must be taken into account for quantitative interpretations of the measured surface density profile. The projection onto the sky plane of a spherically symmetric cluster can be done using the integral:
\begin{equation}
\Sigma(R) = 2 \int_{R}^{\infty}\frac{r n(r)dr}{\sqrt{r^{2}-R^{2}}}
\end{equation}
where $R$ is the distance from the center of the cluster in the plane of the sky, $r$ is the physical radius of the star cluster, $\Sigma(R)$ is the projected surface density, and $n(r)$ is the volume number density. Because this integral diverges for all $\gamma < 1.0$, where $n(r) \propto r^{-\gamma}$, the density in the outer region of the cluster must fall off steeply to accommodate a shallow power law at the center of the cluster. While we do not observe a break in the cluster density profile within our $4\arcsec$ survey, one was observed by \citet{2007A&A...469..125S} and others at scales larger than our survey. In order to account for the outer region of the star cluster, we therefore model the number density profile as a broken power law:
\begin{equation}
n(r) \propto \left\{
	\begin{array}{cc}
	r^{-\gamma_{1}} & r < r_{break}, \\
    r^{-\gamma_{2}} & r \ge r_{break}. 
     \end{array}
     \right.
\end{equation}
Using the measurements from \citet{2007A&A...469..125S}, we set $r_{break} = 8\arcsec$ and $\gamma_{2}=1.8$ (the deprojected values of $R_{break}=6\arcsec$ and $\Gamma = 0.8$, from the surface density measurements using this broken power law model) to constrain the parameters of the outer cluster density profile. We assumed that contamination from young stars is less severe in the region outside of our survey, so using an outer cluster density profile measured from total number counts should introduce only a small bias. For example, the density of bright young stars falls off relatively steeply as $R^{-2}$ \citep[][but see also \citet{2009arXiv0903.2135B}]{2006ApJ...643.1011P,2009ApJ...690.1463L}. To determine the constraints on the physical density profile in the inner region, we performed Monte Carlo simulations of clusters with $\gamma_{1}$ between -2.0 and 2.0. For each value of $\gamma_{1}$, $10^{4}$ realizations were performed. In each realization, 74 stars (the number of late-type stars used to measure the radial profile) were drawn from the broken power law distribution after correcting for the limited area observed in this survey. The locations of these stars were then binned and fitted the same way as the data. We find that for $-2.0 < \gamma_{1} < 0.5$, the projected inner radial profiles are flat with $\Gamma \approx 0$ (see Figure \ref{fig:deproject}). Figure \ref{fig:simcluster} shows the relationship between $\gamma_{1}$ and $\Gamma$ as determined from the MC simulations. Using this relationship and the observed inner radial profile, $\Gamma = -0.27\pm0.16$, we can set an upper limit of 1.0 to the value of $\gamma_{1}$ the 99.73\% confidence level.

Although the slope measurement in this survey cannot constrain whether there is a `hole' in the distribution of late-type stars or just a very shallow power law within the central $4\arcsec$, the inferred slope is significantly flatter than the range of $\gamma_{1}$ between 7/4 and 3/2 predicted by \citet{1977ApJ...216..883B}. In fact, a flat core density profile without a cusp can fit the data equally well. Below, we discuss several plausible dynamical scenarios that may deplete the number density of the late-type giants observed in this study, as well as possible observational prospects for making further progress.

\subsection{Mass segregation}
The true population of dark mass that consists of stellar remnants such as neutron stars or stellar black holes is unknown, but theoretically, there is a strong case for a large population of such remnants due to mass segregation \citep{1993ApJ...408..496M}. As these dark remnants migrate inward, they will scatter the lighter stars outward. Recent simulations by \citet{2008arXiv0808.3150A} showed that a population of $\sim 10$ M$_{\odot}$ stellar remnants can sink into a much steeper density distribution ($2 \lesssim \gamma < 11/4$) than the lighter stars ($3/2 < \gamma < 7/2$), if the stellar remnants are relatively rare compared to stars. This results in a very dense cluster of stellar remnants, but the prediction of the stellar cusp from strong mass segregation does not differ substantially from the \citet{1977ApJ...216..883B} values. Thus, mass segregation cannot be the only mechanism responsible for the inferred lack of a cusp in the evolved red giant population. 

\subsection{Envelope destruction by stellar collisions}

The M and K type giants in this survey have radii that vary between 20 to 200 R$_{\odot}$, which results in substantial cross sections for collisions with other stars in this high density environment \citep[e.g.][]{1999MNRAS.308..257B,2009MNRAS.393.1016D}. This may lead to a preferential depletion of red giants toward the center, which would result in a biased measurement of the underlying stellar cusp slope, since most of the surviving stars in the cusp would be unobserved main sequence stars with smaller radii. The depletion of giants has been suggested in the past based upon low spatial resolution spectroscopy of the CO band-head at 2.29 $\micron$ \citep{1989ApJ...338..824M} as well as in narrow band imaging \citep{2003ApJ...594..812G,2009arXiv0903.2135B}. Theoretically, a collisionally dominated cusp can be as shallow as $\gamma = 0.5$, which is within the range of slopes allowed by our upper limits, for an isotropic distribution, (i.e., for a distribution function that depends only on energy), although this result no longer holds in the case of a more general distribution function \citep{1991ApJ...370...60M}. While the collisional destruction of giants can operate efficiently within the region observed in this survey, the effectiveness of this mechanism drops very quickly with distance from the black hole. For example, \citet{2009MNRAS.393.1016D} found that collisional destruction of giants becomes less likely beyond about 0.1 pc from Sgr A*. While this survey covers a region out to $\sim0.16$ pc in projection from Sgr A*, previous observations such as from \citet{2009arXiv0903.2135B} suggest that the flattening in the late-type giant radial profile extends out even further to $\sim0.24$ pc in projection. If the assumptions about the steep distributions of stellar mass black holes and other stellar remnants in the simulations of \citet{2009MNRAS.393.1016D} are correct, then this would suggest that the collisional destruction of giants is not the dominant mechanism for clearing the cusp of stars. Observational constraints on the effectiveness of this mechanism can be made by comparing the radial density profiles of giants of varying stellar radii. Collisional destruction should preferentially remove stars having larger radii at a given distance from the black hole (see Section \ref{sec:prospects} for further discussion). 

\subsection{IMBH or binary black hole}

A more dramatic change to the stellar cusp can be achieved through the infall of an IMBH. \citet{2006MNRAS.372..174B} performed N body simulations of the effects on the central star cluster with an initial cusp slope of $\gamma = 7/4$ due to the infall of IMBHs of mass $3\times10^{3}$ M$_{\odot}$ and $10^{4}$ M$_{\odot}$. They find that the IMBH will deplete the cusp of stars, which will cause the inner portion of the cluster to resemble a core profile after the IMBH has spiraled into the central black hole, and that it would take over $\sim$ 100 Myr after the IMBH infall to replenish the stellar cusp. For the case of a $10^{4}$ M$_{\odot}$ IMBH, the density profile has a flat core, with a core radius of $\sim 0.2$ pc, which is roughly consistent with that observed in this survey and by \citet{2009arXiv0903.2135B}. 

The existence of a binary massive black hole in the Galactic center can also contribute to the loss of stars in this region. A star passing within the orbit of the companion massive black hole will undergo dynamical interactions with the binary, which can result in the star achieving ejection velocity \citep{2005LRR.....8....8M}. In addition, similar to that of an infalling IMBH, a merger with a massive black hole system more recently than the dynamical relaxation time will result in a system that is still out of equilibrium at this time \citep{2007ApJ...671...53M}. 

Current measurements impose only modest limits on the possibility of an infalling IMBH and a binary black hole. \citet{2008A&A...492..419T} found no obvious kinematic signatures of a recent infall of an IMBH using 3D velocity measurements of the late-type stars, however, a more complete study of the phase space distribution of the stars will be necessary to ascertain the likelihood of an IMBH clearing the cusp of stars (see \citet{2009arXiv0905.4514G} for a more extensive discussion of current constraints on the existence of an IMBH at the Galactic center). Current measurements of the reflex motion of the black hole at radio wavelengths place the limit to the mass of a companion black hole at $< 10^4$ M$_{\odot}$ with semi-major axes $10^{3}$ AU $< a < 10^{5}$ AU \citep{2004ApJ...616..872R}. Precise measurements of the possible reflex motion of the black hole using stellar orbits can provide stronger constraints in the future on the presence of a companion black hole \citep{2008ApJ...689.1044G}.

\subsection{Prospects for future observations}
\label{sec:prospects}

Given the current state of measurements of the cusp density, it is unclear which, if any, of these dynamical interactions are responsible for the shallow inferred cluster slope. The strong degeneracy between the surface number density and the de-projected volume density makes it difficult to determine the extent of the depletion of red giants within the inner 0.1 pc. Limits to the line of sight distance of these stars from Sgr A* will be necessary to resolve this degeneracy. Within the central arcsecond, it may be possible with direct orbit fits to measure the orbital elements of stars located physically near the black hole. For example, at least one of these late-type stars, S0-17 (which passed within 65 mas of Sgr A* in the plane of the sky in 2007) has an orbit measurement that places its periastron distance closer than 8 mpc (1600 AU) from the black hole \citep{2009ApJ...692.1075G}. This indicates that while there are fewer late-type stars expected in the central region, it is not entirely devoid of these stars. Even for stars without well-determined orbits, acceleration limits can be used to constrain the 3D distribution of the late-type population, and thus help determine whether there is a 'hole' or simply a flattening of the radial density profile toward the center of the cluster. 

Determining which mechanism is responsible for the shallow slope of the late-type stars is important because it is related to whether we can use these stars as tracers of the underlying stellar cluster. The main sequence, $\gtrsim 1$ Gyr old stars, represents a significant fraction of both the mass and number density of the stellar cusp, but is currently unobservable spectroscopically; spectral types later than A have $K^{\prime} > 19$, compared to our current sensitivity limit of 15.5 at $K^{\prime}$. Fortunately, in a relaxed population, the red giants should have the same phase space distribution as main sequence stars of comparable mass, so their radial density profile should be identical. However, the red giants have substantially larger radii than the main sequence stars, so they are more likely to be destroyed if collisions are important. It may be possible to observationally constrain whether collisions are the dominant mechanism responsible for the observed slope of the radial density profile by spectral typing the late-type stars more precisely than was possible in this survey. For example, the equivalent width of the CO band head at 2.29 $\micron$, which varies over a wider range than the equivalent width of \ion{Na}{1} observed in this study, should permit one to distinguish between individual spectral sub-types. With a factor of $\sim6$ change in the radii of red giants between spectral types K3III and M5III \citep{1999AJ....117..521V}, a difference in density profile between different spectral types would provide a test of whether collisions affect the distribution of these stars.

\section{Conclusions}

We completed a survey using high angular resolution integral field spectroscopy of the inner $\sim 0.16$ pc of the Galaxy in order to measure the radial profile of the late-type stars in this region. The survey reached a depth of $K^{\prime} < 15.5$ and is able to differentiate between early and late-type stars. The survey provided spectral types for 60 early-type and 74 late-type stars, with 23 and 61 previously unreported early and late-type stars, respectively. We find a late-type stellar surface density power law exponent $\Gamma = -0.27\pm0.19$, which limits the volume number density profile slope $\gamma$ to be less than 1.0 at the 99.73\% confidence level, and even allows for the presence of a central drop in the density of late-type giants. We present three scenarios that may lead to the depletion of late-type stars, but are unable to constrain the candidate mechanism sufficiently to determine which is dominant. Given the current measurements, we cannot yet determine whether the distribution of observed giants is representative of the distribution of stellar mass. Being able to infer the underlying dynamically relaxed stellar population will be crucial in order to establish whether the Galactic center is lacking the type of stellar cusp long predicted by theory. Obtaining an unbiased measurement of the stellar distribution is important because such cusps have considerable impact on the growth of massive black holes as well as on the evolution of nuclear star clusters.  Progress in achieving this goal will be possible with improved kinematics and spectral coverage in order to break the degeneracy in the surface number density profile to better establish the three-dimensional distribution of the stellar cluster.   

The authors thank the staff of the Keck observatory, especially Randy Campbell, Al Conrad, and Jim Lyke, for all their help in obtaining the new observations and Rainer Sch{\"o}del for providing us with an extinction map of the Galactic center region. Support for this work was provided by NSF
grant AST-0406816 and the NSF Science
\& Technology Center for AO, managed by UCSC
(AST-9876783).
The infrared data presented herein were obtained at the W. M. Keck Observatory, which is operated as a scientific partnership among the California Institute of Technology, the University of California and the National Aeronautics and Space Administration. The Observatory was made possible by the generous financial support of the W. M. Keck Foundation. The authors wish to recognize and acknowledge the very significant cultural role that the summit of Mauna Kea has always had within the indigenous Hawaiian community. We are most fortunate to have the opportunity to conduct observations from this mountain.

{\it Facilities:} \facility{Keck:II (OSIRIS)}.

\begin{figure}
\centering
\includegraphics[angle=90,width=6.5in]{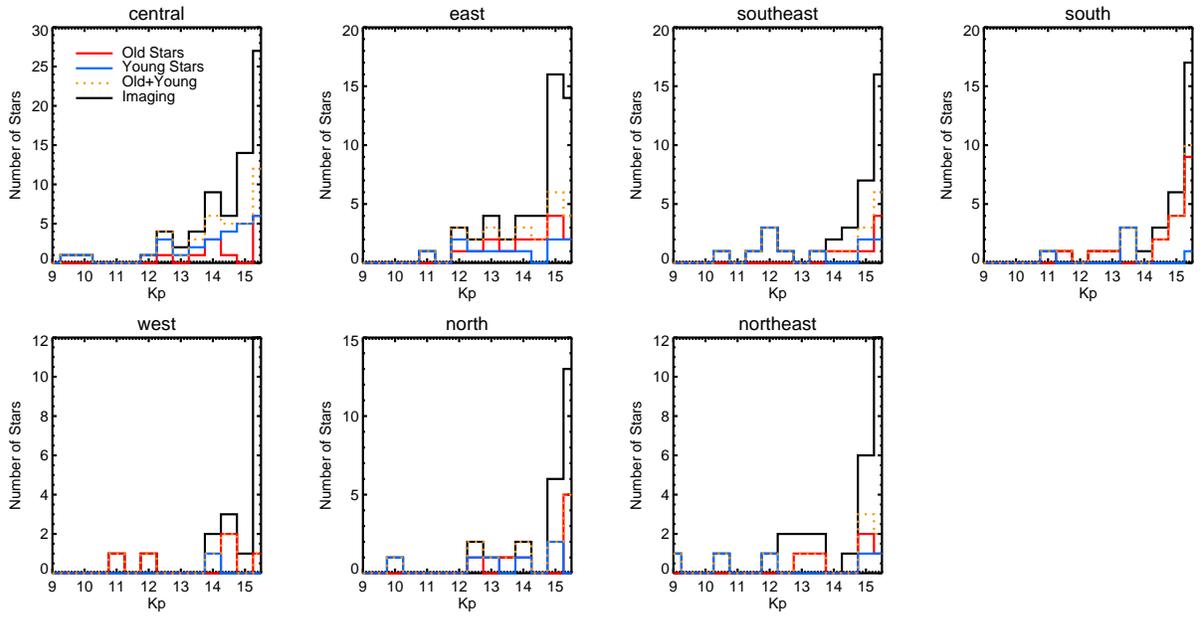}
\caption{The $K^\prime$ luminosity function from the spectroscopic survey in comparison to imaging in each individual pointing. }
\label{fig:kluminosity_fields}
\end{figure}

\clearpage
\begin{deluxetable}{lccccccc}
\tablecolumns{8}
\tablecaption{Summary of OSIRIS observations}
\tablewidth{0pc}
\tabletypesize{\scriptsize}  
\tablehead{\colhead{Field Name} & \colhead{Field Center\tablenotemark{a}} & \colhead{Date} & \colhead{N$_{frames}$\tablenotemark{b}} &  \colhead{FHWM\tablenotemark{c}}  \\
\colhead{} & \colhead{(\arcsec)} & \colhead{(UT)} & \colhead{}  & \colhead{(mas)} }
\startdata
GC Central (C)    & 0, 0            & 2008 May 16 & 11 & $84\times85$  \\ 
GC East (E)      & 2.88, -0.67     & 2007 July 18 & 10 & $85\times70$  \\ 
GC South (S)   & -0.69, -2.00    & 2007 July 19 & 10 & $73\times63$    \\ 
GC West  (W)     & -2.70, 0.74     & 2007 July 20 & 11 & $110\times86$ \\ 
GC Southeast (SE)  & 1.67, -2.23     & 2008 June 03 & 11 & $68\times63$ \\
GC North  (N)    & 0.33, 2.01      & 2008 June 07 & 7  & $102\times85$  \\
              &      & 2008 June 10 & 5  & $75\times70$                 \\
GC Northeast (NE)  & 2.55, 1.27      & 2008 June 10 & 5  & $74\times68$  \\
\enddata
\tablenotetext{a}{RA and DEC offset from Sgr A* (RA offset is positive to the east).}
\tablenotetext{b}{Frames have an integration time of 900 s.}
\tablenotetext{c}{Average FWHM of a relatively isolated star for the night, found from a two-dimensional Gaussian fit to the source.}
\label{table:obs}
\end{deluxetable}

\begin{deluxetable}{ccccccccc}
   \tablecolumns{9}
   \tablewidth{0pc}
   \centering
   \tabletypesize{\scriptsize}  
   \tablecaption{Results and survey completeness} 
   \tablehead{\colhead{} & \colhead{} & \colhead{} & \colhead{} & \colhead{} &
   \multicolumn{2}{c}{Observed} & \multicolumn{2}{c}{$A_{K}$ Corrected\tablenotemark{d}} \\
   \cline{6-7} \cline{8-9} \\
   Field  & \colhead{N$_{SF}$\tablenotemark{a}} & \colhead{N$_{ID}$\tablenotemark{b}}  & \colhead{$A_{k}$\tablenotemark{c}}  & RMS$_{A_{k}}$ & K$^{\prime}$\tablenotemark{e} & Completeness & K$^{\prime}$  & Completeness}
\startdata
C     & 100 & 28   &  3.24   &  0.04 &  15.5 - 16.0  & 0.4 &  15.5 - 16.0 & 0.4 \\
E     & 37 & 25    &  3.05   &  0.06 &  15.5 - 16.0  & 0.3 &  15.0 - 15.5 & 0.3 \\
SE    & 60 & 25    &  3.29   &  0.10 &  15.5 - 16.0  & 0.4 &  15.0 - 15.5 & 0.5 \\
S     & 35 & 6     &  3.28   &  0.05 &  15.5 - 16.0  & 0.6 &  15.0 - 15.5 & 0.6 \\
W     & 35 & 19    &  3.44   &  0.12 &  14.5 - 15.0  & 0.7 &  14.0 - 14.5 & 0.4 \\
N     & 40 & 14    &  3.39   &  0.12 &  15.5 - 16.0  & 0.4 &  15.0 - 15.5 & 0.4 \\
NE    & 32 & 7     &  3.13   &  0.07 &  15.0 - 15.5  & 0.5 &  15.0 - 15.5 & 0.5 \\
\enddata
\tablenotetext{a}{Number of stars detected by \textit{StarFinder}.}
\tablenotetext{b}{Number of stars with determined spectral types.}
\tablenotetext{c}{Extinction values from \citet{2009arXiv0903.2135B}.}
\tablenotetext{d}{Extinction correction using ($A_{k}$ - 3.0), with $A_{k}$ from the extinction map in \citet{2009arXiv0903.2135B}.}
\tablenotetext{e}{Faintest magnitude bin where the completeness is at least 30\%.}
   \label{tab:ext}
\end{deluxetable}

\begin{deluxetable}{lccccccccccc}
\tablecolumns{12}
\tablecaption{OSIRIS observations of late-type stars}
\tablewidth{0pc}
\tabletypesize{\scriptsize}  
\tablehead{\colhead{Name} & \colhead{K$^\prime$} & \colhead{RA offset} & \colhead{DEC offset} & \colhead{R} & \colhead{Epoch} & \colhead{Note\tablenotemark{a}} & \colhead{Na}  & \colhead{$\sigma_{Na}$} & $N_{obs}\tablenotemark{b}$ & SNR\tablenotemark{c} & \colhead{Field} \\
\colhead{} & \colhead{} & \colhead{(\arcsec)} & \colhead{(\arcsec)} & \colhead{(\arcsec)} & \colhead{}  &
		     \colhead{} & \colhead{(\AA)} & \colhead{(\AA)} & \colhead{} & \colhead{} & \colhead{}}
\startdata
\input{tabledata}
\enddata
\tablenotetext{a}{1. Spectral type first reported in this work. Recent spectral identification: 2. \citet{2006ApJ...643.1011P} 3. \citet{2000MNRAS.317..348G} 4. \citet{2009ApJ...692.1075G}.} 
\tablenotetext{b}{Each observation has an integration time of 900 s.}
\tablenotetext{c}{Signal to noise per pixel calculated between 2.212 to 2.218 \micron.}
\label{table:late}
\end{deluxetable}

\begin{deluxetable}{lccccccccccc}
\tablecolumns{12}
\tablecaption{OSIRIS observations of early-type stars}
\tablewidth{0pc}
\tabletypesize{\scriptsize}  
\tablehead{\colhead{Name} & \colhead{K$^\prime$} & \colhead{RA offset} & \colhead{DEC offset} & \colhead{R} & \colhead{Epoch} & \colhead{Note\tablenotemark{a}} & \colhead{Br $\gamma$\tablenotemark{b}}  & \colhead{$\sigma_{Br \gamma}$} & $N_{obs}\tablenotemark{c}$ & SNR\tablenotemark{d} & \colhead{Field}  \\
\colhead{} & \colhead{} & \colhead{(\arcsec)} & \colhead{(\arcsec)} & \colhead{(\arcsec)} & \colhead{} & \colhead{} &
		     \colhead{(\AA)} & \colhead{(\AA)} & \colhead{} & \colhead{} & \colhead{}  }
\startdata
\input{youngtable}
\enddata
\tablenotetext{a}{1. Spectral type first reported in this work. Recent spectral identification: 2. \citet{2006ApJ...643.1011P} 3. \citet{2000MNRAS.317..348G} 4. \citet{2009ApJ...692.1075G}.} 
\tablenotetext{b}{Each observation has an integration time of 900 s.}
\tablenotetext{c}{Stars with no reported equivalent widths have no detectable Br $\gamma$ lines (see Section \ref{sec:spec}), with the exception of the IRS 16 sources, which all have strong Br $\gamma$ emission lines from strong winds. }
\tablenotetext{d}{Signal to noise per pixel calculated between 2.212 to 2.218 \micron.}
\label{table:early}
\end{deluxetable}

\begin{deluxetable}{lcccccc}
\tablecolumns{7}
\tablecaption{OSIRIS spectra with unknown spectral type}
\tablewidth{0pc}
\tabletypesize{\scriptsize}  
\tablehead{\colhead{Name} & \colhead{K$^\prime$} & \colhead{RA offset} & \colhead{DEC offset} & \colhead{R} & \colhead{Epoch} & \colhead{SNR\tablenotemark{a}}  \\
\colhead{} & \colhead{} & \colhead{(\arcsec)} & \colhead{(\arcsec)} & \colhead{(\arcsec)} & \colhead{} & \colhead{}}
\startdata
\input{unknowntable}
\enddata
\tablenotetext{a}{Signal to noise per pixel calculated between 2.212 to 2.218 \micron.}
\label{table:unknown}
\end{deluxetable}

\end{document}

%% file: tabledata.tex
S0-17	 &    15.7	 &    0.02	 &    -0.09	 &    0.1	 &    2004.47	 &    4	 &    1.3	 &    0.17	 &    11	 &    20	 &    C \\ 
S0-6	 &    14.2	 &    0.05	 &    -0.38	 &    0.39	 &    2002.69	 &    4	 &    2.92	 &    0.83	 &    9	 &    44	 &    C \\ 
S0-27	 &    15.7	 &    0.14	 &    0.53	 &    0.54	 &    2003.34	 &    1	 &    2.71	 &    1.69	 &    7	 &    16	 &    C \\ 
S0-29	 &    15.6	 &    0.32	 &    -0.46	 &    0.56	 &    2004.84	 &    4	 &    4.63	 &    1.4	 &    9	 &    6	 &    C \\ 
S0-28	 &    15.8	 &    -0.17	 &    -0.55	 &    0.57	 &    2003.53	 &    4	 &    1.59	 &    1.19	 &    10	 &    35	 &    C \\ 
S0-12	 &    14.4	 &    -0.56	 &    0.39	 &    0.68	 &    2002.69	 &    4	 &    2.6	 &    0.51	 &    4	 &    33	 &    C \\ 
S0-13	 &    13.5	 &    0.54	 &    -0.44	 &    0.69	 &    2002.69	 &    4	 &    2.8	 &    0.24	 &    6	 &    77	 &    C \\ 
S1-5	 &    12.8	 &    0.34	 &    -0.92	 &    0.98	 &    2002.69	 &    4	 &    3.22	 &    0.64	 &    4	 &    45	 &    C \\ 
S1-10	 &    14.9	 &    -1.13	 &    -0.03	 &    1.13	 &    2002.99	 &    4	 &    2.5	 &    0.86	 &    5	 &    25	 &    C \\ 
S1-31	 &    15.7	 &    -0.98	 &    0.56	 &    1.13	 &    2003.6	 &    1	 &    4.62	 &    1.53	 &    4	 &    11	 &    C \\ 
S1-6	 &    15.5	 &    -0.93	 &    0.73	 &    1.18	 &    2003.34	 &    1	 &    3.54	 &    1.01	 &    2	 &    18	 &    C \\ 
S1-36	 &    15.8	 &    -0.63	 &    -1.2	 &    1.35	 &    2003.58	 &    1	 &    2.31	 &    0.77	 &    4	 &    18	 &    S \\ 
S1-13	 &    14.2	 &    -1.12	 &    -0.92	 &    1.45	 &    2002.91	 &    4	 &    2.38	 &    1.5	 &    3	 &    21	 &    C \\ 
S1-15	 &    14.2	 &    -1.36	 &    0.52	 &    1.45	 &    2002.69	 &    1	 &    3.42	 &    0.27	 &    4	 &    38	 &    C \\ 
S1-39	 &    15.4	 &    -0.55	 &    -1.37	 &    1.48	 &    2003.22	 &    1	 &    3.94	 &    0.66	 &    7	 &    16	 &    S \\ 
S1-17	 &    12.5	 &    0.47	 &    -1.49	 &    1.56	 &    2002.69	 &    1	 &    4.33	 &    0.17	 &    4	 &    85	 &    S \\ 
S1-44	 &    15.6	 &    0.3	 &    1.61	 &    1.63	 &    2003.48	 &    1	 &    4.03	 &    0.4	 &    5	 &    22	 &    N \\ 
S1-48	 &    15.5	 &    -0.62	 &    -1.53	 &    1.66	 &    2004.2	 &    1	 &    3.23	 &    0.87	 &    10	 &    32	 &    S \\ 
S1-20	 &    12.7	 &    0.41	 &    1.61	 &    1.66	 &    2002.69	 &    1	 &    4.87	 &    0.12	 &    12	 &    57	 &    N \\ 
S1-23	 &    11.9	 &    -0.93	 &    -1.48	 &    1.74	 &    2002.69	 &    1	 &    3.51	 &    0.1	 &    10	 &    77	 &    S \\ 
S1-25	 &    13.5	 &    1.65	 &    -0.63	 &    1.76	 &    2002.69	 &    3	 &    3.51	 &    	 &    1	 &    33	 &    E \\ 
S1-66	 &    15.6	 &    -0.81	 &    -1.76	 &    1.94	 &    2004.87	 &    1	 &    1.38	 &    0.37	 &    10	 &    8	 &    S \\ 
S1-67	 &    15.6	 &    -1.44	 &    -1.32	 &    1.95	 &    2003.56	 &    1	 &    2.61	 &    0.72	 &    9	 &    30	 &    S \\ 
S1-68	 &    13.4	 &    1.86	 &    -0.61	 &    1.96	 &    2002.69	 &    1	 &    2.47	 &    0.52	 &    4	 &    50	 &    E \\ 
S2-12	 &    15.3	 &    1.67	 &    1.17	 &    2.04	 &    2003.08	 &    1	 &    4.41	 &    0.16	 &    5	 &    18	 &    NE \\ 
S2-34	 &    15.4	 &    1.81	 &    0.97	 &    2.05	 &    2003.43	 &    1	 &    3.8	 &    1	 &    5	 &    18	 &    NE \\ 
S2-11	 &    12	 &    1.97	 &    -0.62	 &    2.07	 &    2002.69	 &    1	 &    1.33	 &    0.33	 &    8	 &    98	 &    E \\ 
S2-3	 &    14.5	 &    -1.54	 &    -1.41	 &    2.09	 &    2002.84	 &    1	 &    2.92	 &    0.23	 &    9	 &    61	 &    S \\ 
S2-37	 &    15.8	 &    0.08	 &    2.1	 &    2.11	 &    2003.91	 &    1	 &    3.14	 &    0.55	 &    12	 &    17	 &    N \\ 
S2-2	 &    14.1	 &    -0.55	 &    2.06	 &    2.13	 &    2002.88	 &    1	 &    2.26	 &    0.21	 &    11	 &    42	 &    N \\ 
S2-41	 &    15.5	 &    -0.47	 &    -2.1	 &    2.16	 &    2004.41	 &    1	 &    2.86	 &    1.3	 &    10	 &    36	 &    S \\ 
S2-8	 &    12.2	 &    -1.84	 &    0.96	 &    2.08	 &    2002.69	 &    1	 &    2.82	 &    0.66	 &    9	 &    53	 &    W \\ 
S2-134	 &    15.6	 &    -0.99	 &    -1.99	 &    2.23	 &    2005.79	 &    1	 &    3.66	 &    0.84	 &    10	 &    8	 &    S \\ 
S2-49	 &    15.5	 &    -0.83	 &    -2.1	 &    2.26	 &    2004.14	 &    1	 &    2.88	 &    0.46	 &    10	 &    19	 &    S \\ 
S2-47	 &    14.2	 &    2.19	 &    -0.53	 &    2.25	 &    2002.75	 &    1	 &    2.84	 &    0.79	 &    10	 &    41	 &    E \\ 
S2-51	 &    15.9	 &    0.76	 &    2.18	 &    2.31	 &    2004.88	 &    1	 &    4.88	 &    2.19	 &    12	 &    23	 &    N \\ 
S2-18	 &    13.3	 &    -0.98	 &    -2.14	 &    2.35	 &    2002.83	 &    3	 &    3.18	 &    0.17	 &    10	 &    57	 &    S \\ 
S2-23	 &    14.8	 &    1.66	 &    1.77	 &    2.42	 &    2002.93	 &    1	 &    3.64	 &    0.38	 &    4	 &    36	 &    NE \\ 
S2-57	 &    14.7	 &    -1.21	 &    -2.12	 &    2.44	 &    2003.76	 &    1	 &    3.06	 &    0.3	 &    10	 &    45	 &    S \\ 
S2-26	 &    13.7	 &    0.74	 &    2.42	 &    2.53	 &    2002.65	 &    3	 &    3.65	 &    0.34	 &    12	 &    43	 &    N \\ 
S2-62	 &    15.2	 &    -1.07	 &    -2.32	 &    2.56	 &    2004.67	 &    1	 &    2.89	 &    0.37	 &    10	 &    31	 &    S \\ 
S2-25	 &    14.1	 &    0.74	 &    -2.45	 &    2.56	 &    2003.5	 &    1	 &    3.54	 &    0.21	 &    8	 &    91	 &    SE \\ 
S2-205	 &    15.8	 &    0.48	 &    2.51	 &    2.56	 &    2005.54	 &    1	 &    3.34	 &    1.31	 &    12	 &    22	 &    N \\ 
S2-65	 &    15.8	 &    2.35	 &    -1.06	 &    2.58	 &    2004.13	 &    1	 &    1.96	 &    0.22	 &    10	 &    18	 &    E \\ 
S2-212	 &    15.7	 &    0.18	 &    -2.59	 &    2.59	 &    2006.75	 &    1	 &    3.31	 &    2.01	 &    7	 &    12	 &    S \\ 
S2-70	 &    14.5	 &    -2.64	 &    0.41	 &    2.67	 &    2003.98	 &    1	 &    3.74	 &    0.9	 &    11	 &    29	 &    W \\ 
S2-71	 &    15.3	 &    -0.9	 &    -2.53	 &    2.68	 &    2004.72	 &    1	 &    1.99	 &    0.16	 &    10	 &    30	 &    S \\ 
S2-72	 &    15.1	 &    -1.46	 &    -2.28	 &    2.71	 &    2004.32	 &    1	 &    3.98	 &    0.51	 &    10	 &    32	 &    S \\ 
S2-73	 &    15.2	 &    2.16	 &    -1.64	 &    2.72	 &    2003.15	 &    1	 &    3.09	 &    0.12	 &    11	 &    49	 &    SE \\ 
S2-75	 &    14.5	 &    2.63	 &    -0.89	 &    2.78	 &    2002.69	 &    1	 &    2.49	 &    0.7	 &    10	 &    39	 &    E \\ 
S2-31	 &    13.1	 &    2.81	 &    -0.2	 &    2.81	 &    2002.69	 &    1	 &    2.96	 &    0.12	 &    10	 &    71	 &    E \\ 
S2-80	 &    15.6	 &    2.23	 &    1.78	 &    2.85	 &    2003.58	 &    1	 &    2.4	 &    0.91	 &    5	 &    20	 &    NE \\ 
S2-81	 &    15.6	 &    1.97	 &    -2.07	 &    2.86	 &    2004.75	 &    1	 &    2.98	 &    0.53	 &    11	 &    22	 &    SE \\ 
S2-83	 &    15.6	 &    2.86	 &    -0.69	 &    2.94	 &    2003.62	 &    1	 &    2.17	 &    1.7	 &    10	 &    21	 &    E \\ 
S2-319	 &    15.7	 &    0.99	 &    -2.79	 &    2.96	 &    2005.71	 &    1	 &    2.62	 &    0.81	 &    8	 &    14	 &    SE \\ 
S2-32	 &    12.4	 &    1.12	 &    2.78	 &    2.99	 &    2003.05	 &    1	 &    3.53	 &    0.15	 &    6	 &    59	 &    N \\ 
S3-68	 &    15.8	 &    -0.35	 &    3.06	 &    3.08	 &    2005.93	 &    1	 &    3.96	 &    0.53	 &    5	 &    21	 &    N \\ 
S3-14	 &    13.9	 &    0.1	 &    3.09	 &    3.09	 &    2004.54	 &    1	 &    1.98	 &    	 &    1	 &    18	 &    N \\ 
S3-16	 &    15.3	 &    2.99	 &    -0.95	 &    3.13	 &    2003.76	 &    1	 &    5.3	 &    1.3	 &    10	 &    24	 &    E \\ 
S3-20	 &    14.9	 &    1.58	 &    -2.8	 &    3.21	 &    2003.83	 &    1	 &    1.64	 &    0.53	 &    11	 &    54	 &    SE \\ 
S3-6	 &    12.9	 &    3.22	 &    0.01	 &    3.22	 &    2002.8	 &    1	 &    3.36	 &    0.26	 &    10	 &    77	 &    E \\ 
S3-109	 &    15.6	 &    -3.2	 &    0.43	 &    3.23	 &    2006.75	 &    1	 &    4.32	 &    1.96	 &    11	 &    20	 &    W \\ 
S3-151	 &    15.7	 &    1.9	 &    -2.78	 &    3.37	 &    2006.22	 &    1	 &    2.99	 &    1.56	 &    11	 &    16	 &    SE \\ 
S3-32	 &    15.5	 &    2.91	 &    -1.77	 &    3.41	 &    2004.51	 &    1	 &    3.24	 &    0.62	 &    9	 &    37	 &    SE \\ 
S3-8	 &    14	 &    3.41	 &    -0.49	 &    3.44	 &    2002.81	 &    1	 &    2.43	 &    0.67	 &    10	 &    56	 &    E \\ 
S3-11	 &    15.1	 &    2.96	 &    -1.93	 &    3.53	 &    2004.12	 &    3	 &    4.05	 &    0.84	 &    4	 &    35	 &    SE \\ 
S3-36	 &    14.7	 &    3.46	 &    -0.83	 &    3.55	 &    2003.76	 &    1	 &    3.65	 &    0.11	 &    9	 &    57	 &    E \\ 
S3-249	 &    14.7	 &    -3.38	 &    1.32	 &    3.63	 &    2006.75	 &    1	 &    3.3	 &    0.34	 &    11	 &    33	 &    W \\ 
S3-37	 &    15.1	 &    3.39	 &    1.35	 &    3.65	 &    2004.66	 &    1	 &    3.45	 &    1.23	 &    5	 &    17	 &    NE \\ 
S3-38	 &    15	 &    3.69	 &    -0.11	 &    3.69	 &    2005.34	 &    1	 &    2.76	 &    0.92	 &    9	 &    33	 &    E \\ 
S3-39	 &    13.6	 &    3.57	 &    1.1	 &    3.73	 &    2002.97	 &    1	 &    3.08	 &    0.53	 &    3	 &    32	 &    NE \\ 
S3-385	 &    15.3	 &    3.67	 &    -1.5	 &    3.96	 &    2005.81	 &    1	 &    1.32	 &    0.36	 &    6	 &    26	 &    E \\ 
S4-1	 &    13.5	 &    4.02	 &    -0.37	 &    4.04	 &    2003.92	 &    1	 &    3.09	 &    0.25	 &    9	 &    102	 &    E \\ 
S4-2	 &    13	 &    3.75	 &    1.67	 &    4.1	 &    2004.01	 &    1	 &    3.51	 &    0.33	 &    3	 &    52	 &    NE \\ 

%% file: youngtable.tex
S0-2	 &    14.1	 &    0	 &    0.16	 &    0.16	 &    2005.61	 &    2	 &    5.4	 &    0.4	 &    9	 &    57	 &    C \\ 
S0-1	 &    14.7	 &    0.04	 &    -0.26	 &    0.27	 &    2006.3	 &    2	 &    0.5	 &    0.3	 &    9	 &    31	 &    C \\ 
S0-19	 &    15.5	 &    -0.04	 &    0.35	 &    0.36	 &    2006.25	 &    2	 &    5.7	 &    2.3	 &    11	 &    16	 &    C \\ 
S0-3	 &    14.5	 &    0.31	 &    0.12	 &    0.33	 &    2006.38	 &    2	 &    5.4	 &    3.6	 &    9	 &    21	 &    C \\ 
S0-5	 &    15.1	 &    0.18	 &    -0.36	 &    0.4	 &    2006.32	 &    2	 &    7.8	 &    1.1	 &    9	 &    15	 &    C \\ 
S0-4	 &    14.5	 &    0.4	 &    -0.28	 &    0.49	 &    2005.83	 &    2	 &    5.4	 &    6	 &    10	 &    1	 &    C \\ 
S0-7	 &    15.5	 &    0.49	 &    0.1	 &    0.5	 &    2006.36	 &    2	 &    6.5	 &    2	 &    10	 &    12	 &    C \\ 
S0-11	 &    15.3	 &    0.5	 &    -0.05	 &    0.5	 &    2006.73	 &    2	 &    10.2	 &    7.6	 &    9	 &    21	 &    C \\ 
S0-9	 &    14.4	 &    0.19	 &    -0.58	 &    0.61	 &    2006.74	 &    4	 &    5.4	 &    2	 &    7	 &    38	 &    C \\ 
S0-31	 &    15.1	 &    0.54	 &    0.44	 &    0.7	 &    2006.47	 &    2	 &    6.01	 &    1.54	 &    6	 &    21	 &    C \\ 
S0-14	 &    13.7	 &    -0.77	 &    -0.27	 &    0.82	 &    2005.51	 &    2	 &    4.1	 &    0.7	 &    7	 &    76	 &    C \\ 
S0-15	 &    13.7	 &    -0.96	 &    0.24	 &    0.99	 &    2005.8	 &    2	 &    2.4	 &    0.5	 &    7	 &    67	 &    C \\ 
S1-3	 &    12.1	 &    0.4	 &    0.88	 &    0.96	 &    2004.98	 &    2	 &    	 &    	 &    5	 &    79	 &    C \\ 
S1-1	 &    13.1	 &    1.01	 &    0.03	 &    1.01	 &    2005.2	 &    4	 &    1.72	 &    1	 &    5	 &    57	 &    C \\ 
S1-2	 &    14.9	 &    0.01	 &    -1.01	 &    1.01	 &    2005.93	 &    2	 &    4.7	 &    3.5	 &    5	 &    19	 &    C \\ 
S1-4	 &    12.6	 &    0.83	 &    -0.67	 &    1.07	 &    2005.84	 &    1	 &    	 &    	 &    5	 &    52	 &    C \\ 
S1-8	 &    14.2	 &    -0.63	 &    -0.89	 &    1.09	 &    2006.17	 &    2	 &    5.1	 &    2.7	 &    4	 &    42	 &    C \\ 
IRS16NW	 &    10.1	 &    0.05	 &    1.22	 &    1.22	 &    2004.74	 &    2 \\ 
IRS16C	 &    9.9	 &    1.1	 &    0.51	 &    1.21	 &    2003.89	 &    2 \\ 
S1-33	 &    15	 &    -1.25	 &    -0.01	 &    1.25	 &    2006.26	 &    1	 &    5.2	 &    1.6	 &    5	 &    17	 &    C \\ 
S1-34	 &    13.2	 &    0.85	 &    -0.99	 &    1.31	 &    2006.38	 &    1	 &    	 &    	 &    3	 &    52	 &    C \\ 
S1-12	 &    13.8	 &    -0.8	 &    -1.01	 &    1.29	 &    2006.21	 &    2	 &    3.7	 &    1.5	 &    4	 &    40	 &    C \\ 
S1-14	 &    12.8	 &    -1.34	 &    -0.33	 &    1.38	 &    2005.49	 &    1	 &    0.5	 &    0.27	 &    3	 &    52	 &    C \\ 
IRS16SW	 &    10.1	 &    1.07	 &    -0.96	 &    1.44	 &    2004.89	 &    2 \\ 
S1-19	 &    13.8	 &    0.38	 &    -1.62	 &    1.66	 &    2005.45	 &    1	 &    7	 &    1	 &    4	 &    37	 &    S \\ 
S1-18	 &    15	 &    -0.76	 &    1.51	 &    1.69	 &    2006.56	 &    1	 &    1.81	 &    1	 &    2	 &    24	 &    N \\ 
S1-21	 &    13.3	 &    -1.66	 &    0.13	 &    1.66	 &    2006.16	 &    2	 &    	 &    	 &    5	 &    47	 &    W \\ 
S1-22	 &    12.7	 &    -1.61	 &    -0.5	 &    1.68	 &    2005.65	 &    2	 &    -1.61	 &    0.49	 &    3	 &    61	 &    C \\ 
S1-24	 &    11.6	 &    0.71	 &    -1.62	 &    1.77	 &    2005.64	 &    2	 &    2.84	 &    	 &    1	 &    74	 &    SE \\ 
S2-5	 &    13.3	 &    1.91	 &    -0.81	 &    2.07	 &    2005.68	 &    1	 &    	 &    	 &    7	 &    58	 &    E \\ 
IRS16CC	 &    10.6	 &    1.99	 &    0.56	 &    2.07	 &    2004.57	 &    2 \\ 
S2-4	 &    12.3	 &    1.47	 &    -1.47	 &    2.08	 &    2006.04	 &    2	 &    1.94	 &    0.19	 &    11	 &    149	 &    SE \\ 
S2-6	 &    12	 &    1.61	 &    -1.35	 &    2.1	 &    2004.69	 &    2	 &    2.04	 &    0.16	 &    7	 &    103	 &    SE \\ 
S2-7	 &    14.1	 &    0.96	 &    1.84	 &    2.08	 &    2006.15	 &    2	 &    0	 &    	 &    12	 &    54	 &    N \\ 
IRS16SW-E	 &    11	 &    1.86	 &    -1.14	 &    2.18	 &    2005.52	 &    2 \\ 
IRS33N	 &    11.4	 &    -0.05	 &    -2.21	 &    2.21	 &    2006.19	 &    2	 &    3.1	 &    0.4	 &    10	 &    104	 &    S \\ 
S2-17	 &    10.9	 &    1.29	 &    -1.88	 &    2.28	 &    2005.14	 &    2	 &    3.19	 &    0.07	 &    11	 &    125	 &    SE \\ 
S2-50	 &    15.5	 &    1.68	 &    -1.52	 &    2.26	 &    2005.38	 &    1	 &    8	 &    3	 &    11	 &    16	 &    SE \\ 
S2-22	 &    12.9	 &    2.3	 &    -0.24	 &    2.31	 &    2005.33	 &    1	 &    -0.73	 &    0.5	 &    10	 &    78	 &    E \\ 
S2-19	 &    12.6	 &    0.42	 &    2.31	 &    2.34	 &    2005.53	 &    2	 &    2.04	 &    0.5	 &    12	 &    68	 &    N \\ 
S2-21	 &    13.5	 &    -1.66	 &    -1.64	 &    2.34	 &    2006.34	 &    1	 &    1.2	 &    0.7	 &    8	 &    71	 &    S \\ 
S2-55	 &    15.4	 &    0.88	 &    -2.19	 &    2.36	 &    2006.53	 &    1	 &    3.2	 &    1.2	 &    11	 &    29	 &    SE \\ 
S2-58	 &    14.1	 &    2.14	 &    -1.16	 &    2.43	 &    2005.74	 &    1	 &    4.7	 &    3	 &    7	 &    41	 &    E \\ 
S2-59	 &    15.4	 &    0.79	 &    -2.34	 &    2.46	 &    2006.48	 &    1	 &    5.6	 &    2	 &    10	 &    25	 &    SE \\ 
S2-61	 &    15.5	 &    2.37	 &    -0.67	 &    2.47	 &    2006.56	 &    1	 &    6	 &    3	 &    10	 &    16	 &    E \\ 
S2-74	 &    13.3	 &    0.15	 &    2.78	 &    2.78	 &    2006.16	 &    2	 &    0	 &    	 &    12	 &    81	 &    N \\ 
S2-77	 &    13.8	 &    -1.78	 &    -2.19	 &    2.82	 &    2006.16	 &    1	 &    1.4	 &    0.5	 &    9	 &    54	 &    S \\ 
S2-76	 &    15.2	 &    -0.23	 &    2.81	 &    2.82	 &    2006.76	 &    1	 &    1.18	 &    1	 &    12	 &    38	 &    N \\ 
S2-317	 &    15.6	 &    -0.74	 &    -2.86	 &    2.95	 &    2006.4	 &    1	 &    5.1	 &    1	 &    3	 &    21	 &    S \\ 
IRS16NE	 &    9.1	 &    2.87	 &    1.01	 &    3.04	 &    2003.62	 &    2 \\ 
S3-2	 &    12.1	 &    3.06	 &    0.53	 &    3.11	 &    2005.43	 &    1	 &    0.49	 &    0.04	 &    3	 &    47	 &    NE \\ 
S3-3	 &    15.1	 &    3.07	 &    -0.67	 &    3.14	 &    2006.47	 &    1	 &    1.9	 &    2	 &    10	 &    21	 &    E \\ 
S3-5	 &    12.2	 &    2.94	 &    -1.18	 &    3.16	 &    2005.66	 &    2	 &    	 &    	 &    10	 &    87	 &    E \\ 
S3-7	 &    13.9	 &    1.91	 &    -2.61	 &    3.24	 &    2006.34	 &    1	 &    0	 &    	 &    11	 &    76	 &    SE \\ 
S3-30	 &    12.8	 &    1.64	 &    -2.96	 &    3.38	 &    2006.39	 &    2	 &    1.63	 &    0.44	 &    8	 &    56	 &    SE \\ 
S3-34	 &    13.5	 &    3.22	 &    -1.26	 &    3.46	 &    2006.29	 &    1	 &    2.12	 &    1.28	 &    9	 &    46	 &    E \\ 
S3-190	 &    14.1	 &    -3.18	 &    1.45	 &    3.49	 &    2006.47	 &    1	 &    1.2	 &    1.1	 &    11	 &    55	 &    W \\ 
S3-10	 &    12.4	 &    3.33	 &    -1.14	 &    3.52	 &    2005.99	 &    2	 &    	 &    	 &    10	 &    81	 &    E \\ 
S3-12	 &    12	 &    2.37	 &    -2.73	 &    3.61	 &    2001.39	 &    1	 &    0	 &    	 &    11	 &    114	 &    SE \\ 
S3-314	 &    15.4	 &    3.81	 &    -0.12	 &    3.82	 &    2006.43	 &    1	 &    6.8	 &    2.6	 &    9	 &    26	 &    E \\ 

%% file: unknowntable.tex
S0-35	 &    15.3	 &    0.06	 &    0.97	 &    0.97	 &    2003.77	 &    16 \\ 
S1-26	 &    15.6	 &    -0.92	 &    0.39	 &    1	 &    2003.07	 &    12 \\ 
S1-29	 &    15.4	 &    1.08	 &    0.17	 &    1.09	 &    2005.09	 &    11 \\ 
S1-32	 &    15.3	 &    -0.96	 &    -0.64	 &    1.15	 &    2003.21	 &    20 \\ 
S1-7	 &    15.9	 &    -1.03	 &    -0.53	 &    1.16	 &    2003.86	 &    13 \\ 
S1-85	 &    15.5	 &    0.92	 &    -0.82	 &    1.23	 &    2006.75	 &    4 \\ 
S1-40	 &    15.9	 &    -1.39	 &    -0.62	 &    1.52	 &    2004.23	 &    7 \\ 
S1-41	 &    15.8	 &    0.99	 &    1.15	 &    1.52	 &    2003.69	 &    4 \\ 
S1-42	 &    15.9	 &    0.94	 &    1.26	 &    1.57	 &    2004.04	 &    9 \\ 
S1-45	 &    15.3	 &    -1.27	 &    1.04	 &    1.65	 &    2003.04	 &    13 \\ 
S1-47	 &    15.6	 &    -1.58	 &    0.46	 &    1.65	 &    2003.14	 &    1 \\ 
S1-50	 &    15.4	 &    1.5	 &    0.68	 &    1.65	 &    2003.7	 &    9 \\ 
S1-128	 &    15.8	 &    1.33	 &    -0.99	 &    1.66	 &    2006.55	 &    0 \\ 
S1-51	 &    15	 &    -1.65	 &    -0.18	 &    1.66	 &    2002.99	 &    10 \\ 
S1-52	 &    15.2	 &    0	 &    1.67	 &    1.67	 &    2003.15	 &    28 \\ 
S1-53	 &    15.3	 &    1.67	 &    -0.12	 &    1.67	 &    2003.12	 &    11 \\ 
S1-54	 &    15.5	 &    -1.52	 &    0.74	 &    1.69	 &    2004.25	 &    3 \\ 
S1-55	 &    15.5	 &    1.59	 &    0.6	 &    1.69	 &    2004.17	 &    7 \\ 
S1-56	 &    15.8	 &    -1.11	 &    1.35	 &    1.75	 &    2003.85	 &    11 \\ 
S1-61	 &    15.9	 &    -1.49	 &    -1	 &    1.79	 &    2003.96	 &    6 \\ 
S1-62	 &    15.3	 &    0.5	 &    1.75	 &    1.82	 &    2003.21	 &    19 \\ 
S1-59	 &    15.5	 &    0.02	 &    1.82	 &    1.82	 &    2003.33	 &    27 \\ 
S1-147	 &    15.9	 &    1.18	 &    -1.47	 &    1.89	 &    2006.75	 &    7 \\ 
S1-64	 &    15.5	 &    0.64	 &    1.81	 &    1.92	 &    2003.08	 &    13 \\ 
S1-159	 &    15.9	 &    1.3	 &    -1.43	 &    1.93	 &    2005.94	 &    9 \\ 
S1-65	 &    15.7	 &    1.46	 &    1.27	 &    1.93	 &    2003.85	 &    5 \\ 
S2-89	 &    15.9	 &    1.07	 &    -1.7	 &    2.01	 &    2006.75	 &    5 \\ 
S2-42	 &    15.6	 &    0.46	 &    2.1	 &    2.15	 &    2004.02	 &    23 \\ 
S2-200	 &    15.7	 &    -2.53	 &    0	 &    2.53	 &    2005.25	 &    12 \\ 
S2-63	 &    15.4	 &    -0.65	 &    2.48	 &    2.56	 &    2003.98	 &    12 \\ 
S2-219	 &    15.8	 &    -1.6	 &    -2.07	 &    2.62	 &    2005.34	 &    8 \\ 
S2-261	 &    15.6	 &    -2.61	 &    1	 &    2.8	 &    2006.04	 &    11 \\ 
S2-268	 &    15.5	 &    -2.78	 &    0.49	 &    2.82	 &    2006.32	 &    13 \\ 
S2-304	 &    15.9	 &    2.52	 &    1.47	 &    2.91	 &    2006.39	 &    12 \\ 
S2-306	 &    15.6	 &    -0.48	 &    -2.89	 &    2.93	 &    2006.75	 &    17 \\ 
S2-84	 &    15.6	 &    1.65	 &    -2.47	 &    2.97	 &    2004.29	 &    26 \\ 
S3-48	 &    15.9	 &    -3	 &    0.37	 &    3.02	 &    2006.32	 &    15 \\ 
S3-50	 &    15.7	 &    -1.87	 &    -2.38	 &    3.03	 &    2006.75	 &    13 \\ 
S3-51	 &    15.1	 &    -0.15	 &    3.02	 &    3.03	 &    2005.88	 &    15 \\ 
S3-62	 &    15.7	 &    2.68	 &    -1.48	 &    3.06	 &    2006.4	 &    28 \\ 
S3-65	 &    15.7	 &    -1.27	 &    -2.8	 &    3.07	 &    2005.72	 &    11 \\ 
S3-4	 &    14.7	 &    3.08	 &    -0.52	 &    3.13	 &    2002.91	 &    28 \\ 
S3-90	 &    15.8	 &    2.74	 &    -1.56	 &    3.15	 &    2005.21	 &    30 \\ 
S3-91	 &    15.7	 &    -1.73	 &    -2.67	 &    3.18	 &    2006.75	 &    8 \\ 
S3-116	 &    15.9	 &    2.25	 &    2.37	 &    3.27	 &    2006.11	 &    9 \\ 
S3-24	 &    15.6	 &    3.26	 &    0.41	 &    3.29	 &    2004.87	 &    4 \\ 
S3-167	 &    15.5	 &    -3.08	 &    1.42	 &    3.4	 &    2006.75	 &    16 \\ 
S3-169	 &    15.5	 &    -3.39	 &    0.38	 &    3.41	 &    2006.75	 &    27 \\ 
S3-31	 &    15.5	 &    3.4	 &    0.38	 &    3.42	 &    2004.42	 &    7 \\ 
S3-33	 &    15.4	 &    3.32	 &    -0.85	 &    3.43	 &    2004.25	 &    16 \\ 
S3-286	 &    15.3	 &    3.42	 &    -1.54	 &    3.75	 &    2004.66	 &    13 \\ 
S3-319	 &    15.4	 &    3.55	 &    -1.42	 &    3.83	 &    2006.39	 &    17 \\ 
S4-47	 &    15.9	 &    2.56	 &    -3.23	 &    4.12	 &    2006.75	 &    11 \\ 
S4-67	 &    15.7	 &    2.8	 &    -3.05	 &    4.14	 &    2006.75	 &    7 \\ 